\documentclass[10pt, conference, compsocconf]{IEEEtran}
\IEEEoverridecommandlockouts
\usepackage{epsfig,graphics,subfigure}

\newtheorem{remark}{Remark}

\newtheorem{theorem}{Theorem}
\usepackage{amsfonts}
\usepackage[hyphens]{url}
\usepackage{dsfont}
\usepackage{amsmath,amssymb}
\begin{document}

\title{Can P2P Technology Benefit Eyeball ISPs? A Cooperative Profit Distribution Answer}

\author{
\IEEEauthorblockN{Ke Xu}
\IEEEauthorblockA{Department of Computer Science\\
Tsinghua University, Beijing, P.R.China\\
Email: xuke@tsinghua.edu.cn}
\and
\IEEEauthorblockN{Yifeng Zhong}
\IEEEauthorblockA{Department of Computer Science\\
Tsinghua University, Beijing, P.R.China\\
Email: victorzhyf@gmail.com}
\and
\IEEEauthorblockN{Huan He \thanks{The main part of this work was finished while Huan He was a student at Tsinghua University.}}
\IEEEauthorblockA{Global Technology Service\\
IBM, North China\\
Email: hehuanbj@cn.ibm.com}
}

\maketitle
\begin{abstract}
\mbox{Peer-to-Peer} (P2P) technology has been regarded as a promising way to help Content Providers (CPs) \mbox{cost-effectively} distribute content. However, under the traditional Internet pricing mechanism, the fact that most P2P traffic flows among peers can dramatically decrease the profit of ISPs, who may take actions against P2P and impede the progress of P2P technology. In this paper, we develop a mathematical framework to analyze such economic issues. Inspired by the idea from cooperative game theory, we propose a cooperative \mbox{profit-distribution} model based on Nash Bargaining Solution (NBS), in which eyeball ISPs and Peer-assisted CPs (PCPs) form two coalitions respectively and then compute a fair Pareto point to determine profit distribution. Moreover, we design a fair and feasible mechanism for profit distribution within each coalition. We show that such a cooperative method not only guarantees the fair profit distribution among network participators, but also helps to improve the economic efficiency of the overall network system. To our knowledge, this is the first work that systematically studies solutions for P2P caused unbalanced profit distribution and gives a feasible cooperative method to increase and fairly share profit.
\end{abstract}
\section{Introduction}
\label{introduction}
\mbox{Peer-to-Peer} (P2P) or \mbox{peer-assisted} architecture offers great potential for Content Providers (CPs) to \mbox{cost-effectively} distribute content by capitalizing network resources of end users. Its economical superiority to the traditional \mbox{Client/Server} (C/S) architecture has been demonstrated by lots of academic work \cite{P2P_VOD-design,P2P-vod,P2P-vod-design2,P2P_saving,P2P_saving2} and successful commercial systems. We believe more and more CPs will adopt P2P technology in the future. This trend seems irreversible.\\
\indent However, under traditional pricing mechanism, \mbox{free-riding} P2P traffic can cause unbalanced profit distribution between Peer-assisted CPs (PCPs) and eyeball Internet Service Providers (ISPs) \cite{Revenue-sharing}. As we know, for eyeball ISPs who directly connect end users and CPs, they usually charge users at a flat price \cite{flat,P2Pfea,his_price2}. Then, for users, P2P traffic can ``\mbox{free-ride}'' in each eyeball ISP's network. This will stimulate users to consume more P2P services. Then, PCPs' profit will sharply increase, while eyeball ISPs' profit will dramatically decrease. It should be noted that unlike eyeball ISPs, transit ISPs \cite{Revenue-sharing} often charge their attached eyeball ISPs based on exchanged traffic \cite{95th} and thus do not give the chance of free riding to P2P. Thus, throughout this paper, when referring to ISPs, we always mean eyeball ISPs. \\
\indent The unbalanced profit distribution between PCPs and ISPs will drive ISPs to take actions against \mbox{free-riding} P2P, which include engineering \cite{cat-mouse,P2P_ISP2,P2P-ISP,P4P} and pricing strategies \cite{CP-participation,uplink-pricing,survey}. We only consider the latter because such imbalance is caused by ISPs' flat pricing model. An effective strategy is to change flat pricing model on users' side to a \mbox{volume-based} pricing model \cite{P2Pfea,uplink-pricing,survey}. Then, although ISPs' profit can be guaranteed at a reasonable level, P2P applications will be less attractive to users as a higher fee, causing a sharp decrease in PCPs' profit. \\
\indent Now that the unbalanced profit distribution can finally impede the wide adoption of P2P technology, someone may ask: \emph{Can we find a \mbox{profit-distribution} model in which P2P technology can also benefit ISPs?} In this paper, we give a positive answer to this question.\\
\indent Inspired by the idea from cooperative game theory, we propose a cooperative \mbox{profit-distribution} model based on the concept of Nash bargaining \cite{NBS}. In this model, ISPs and PCPs form two coalitions and cooperate to maximize their total profit by stimulating P2P service and fairly dividing profit. To guarantee stability, we also consider proper mechanism for profit distribution within each coalition. The main contributions of this paper are listed as follows:
\begin{enumerate}
  \item We build a mathematical framework to describe the multilateral interactions among ISPs, CPs and users in three possible \mbox{non-cooperative} states;
  \item We propose a cooperative \mbox{profit-distribution} model in which P2P technology can fairly benefit both PCP and ISP coalitions;
  \item We design a fair and feasible mechanism for profit distribution within each coalition;
  \item We give examples to prove the effectiveness of the cooperative profit-distribution model.
\end{enumerate}
\indent The rest of the paper is organized as follows. In Section~\ref{non-cooperative game}, we discuss the {non-cooperative} pricing interactions among ISPs, PCPs and users. A cooperative profit distribution model is proposed in Section~\ref{newnbs}. Further, we present our mechanism for profit distribution within each coalition in Section~\ref{sharing}. In Section~\ref{related-work}, we discuss the related work followed by our conclusion in Section~\ref{conclude}.
\section{Non-cooperative Game model}
\label{non-cooperative game}

In this section, we first present a network model and explore the \mbox{multi-lateral} economic relationships among network participators using a \mbox{two-stage} game model. Then, we extend our model with P2P providers (PCPs), which results in three possible \mbox{non-cooperative} market states as possible equilibrium.

\subsection{Network Model}
\label{network-model}

\begin{figure}[htp]
\centering
\includegraphics[width=2.5in]{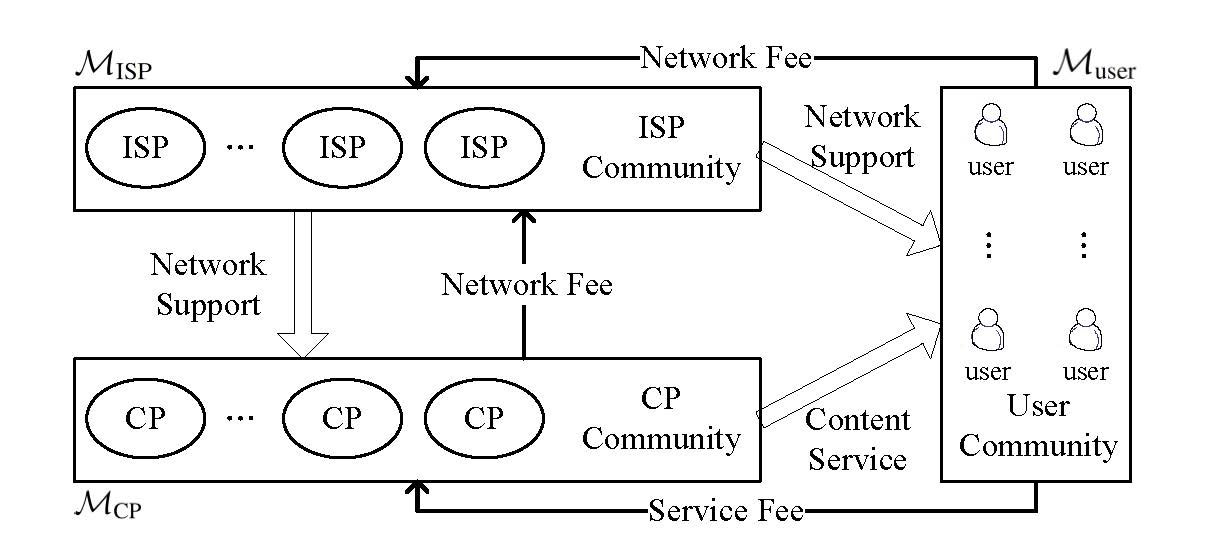}
\caption{Relationships among ISP, CP and user community.}\label{system-relation}
\end{figure}

The network model consists of three communities: ISP community, CP community and user community, denoted by $\mathcal{M}_{\text{ISP}}$, $\mathcal{M}_{\text{CP}}$ and $\mathcal{M}_{\text{user}}$ respectively. Their relationships are shown in Fig.~\ref{system-relation}. In a practical network system, $\mathcal{M}_{\text{ISP}}$ often adopts a \mbox{bandwidth-based} pricing model (such as the \mbox{95-percentile} billing for burstable bandwidth \cite{95th}) to charge $\mathcal{M}_{\text{CP}}$ and a flat pricing model to charge $\mathcal{M}_{\text{user}}$ \cite{his_price2,cp-price}. $\mathcal{M}_{\text{CP}}$ often charges $\mathcal{M}_{\text{user}}$ based on its traffic volume.
\\
\indent In C/S network, all service contents flow from $\mathcal{M}_{\text{CP}}$ to $\mathcal{M}_{\text{user}}$ through $\mathcal{M}_{\text{ISP}}$'s network. Suppose the total traversing bandwidth provided by $\mathcal{M}_{\text{ISP}}$ is $b_{\text{ISP}}$, the bandwidth bought by $\mathcal{M}_{\text{CP}}$ from $\mathcal{M}_{\text{ISP}}$ is $b_{\text{CP}}$ and that bought by $\mathcal{M}_{\text{user}}$ is $b_{\text{user}}$. For $\mathcal{M}_{\text{CP}}$ and $\mathcal{M}_{\text{user}}$, suppose their average bandwidth usage ratios are $\xi_{\text{CP}}$ and $\xi_{\text{user}}$, respectively. Let $v$ be the traffic volume. Then, as traffic is balanced, we have
\begin{equation}\label{v-cs}
v=b_{\text{CP}}\cdot\xi_{\text{CP}}=b_{\text{user}}\cdot\xi_{\text{user}}.
\end{equation}
\indent In \mbox{peer-assisted} network, $\mathcal{M}_{\text{CP}}=\mathcal{M}_{\text{PCP}}\cup\mathcal{M}_{\text{CP}}^r$, where $\mathcal{M}_{\text{PCP}}$ is the set of PCPs and $\mathcal{M}_{\text{CP}}^r$ consists of the rest CPs. We suppose the traffic of $\mathcal{M}_{\text{PCP}}$ accounts for a proportion $\alpha$ in the total traffic of $\mathcal{M}_{\text{CP}}$. Generally, the P2P content provided by servers of PCPs accounts for only a small proportion $\beta$ and the rest $1-\beta$ proportion is provided by $\mathcal{M}_{\text{user}}$. In this case, $\mathcal{M}_{\text{PCP}}$ can reduce its bought bandwidth to a smaller value $b_{\text{PCP}}^*$ to save cost and keep its bandwidth usage ratio at $\xi_{\text{CP}}$; while $\mathcal{M}_{\text{user}}$ with fixed bandwidth at $b_{\text{user}}$ will increase its bandwidth usage ratio to a higher value $\xi_{\text{user}}^*$. Here, we assume the traffic of $\mathcal{M}_{\text{CP}}^r$ (background traffic) will not be impacted by P2P traffic. Then $\mathcal{M}_{\text{CP}}^r$  will keep its traffic at $v_{\text{cs}}=b_{\text{CP}}\cdot (1-\alpha)\cdot \xi_{{\text{CP}}}$. So for P2P content, suppose total \mbox{user-side} uploading and downloading P2P traffic amount is $v_{\text{p2p}}$ and it is not larger than $b_{\text{user}}\cdot \xi_{{\text{CP}}}-v_{\text{cs}}$, we will have
\begin{equation*}
\frac{b_{\text{PCP}}^*\cdot\xi_{\text{CP}}}{\beta}=\frac{v_{\text{p2p}}}{1+(1-\beta)}.
\end{equation*}
Then, similar to the case of C/S network, we should have
\begin{equation}
v_{\text{p2p}}+v_{\text{cs}}=b_{\text{PCP}}^*\cdot\xi_{\text{CP}}\cdot\frac{2-\beta}{\beta}+b_{\text{CP}}\cdot(1-\alpha)\cdot\xi_{{\text{CP}}}= b_{\text{user}}\cdot\xi_{\text{user}}^*,
\end{equation}
where $\xi_{{\text{CP}}}\geq\xi_{\text{user}}^*\geq\xi_{\text{user}}$. Here we assume $\beta>0$, which means server always provides content and makes the equation meaningful.

\subsection{Basic Two-stage Non-cooperative Game}
\label{basic-state0}

We analyze a \mbox{two-stage} game model to determine bandwidth requirement $b_{\text{user}}$ of $\mathcal{M}_{\text{user}}$. We also aim to derive the basic traffic usage $v$ at equilibrium. We use \emph{backward induction} to solve this game and obtain an initial equilibrium market state (State 0).

\subsubsection{Game formulation}

We first form multi-lateral utilities of network participators to explain their relationships shown in Fig.~\ref{system-relation} and then present our \mbox{two-stage} game formulation. At the first stage, $\mathcal{M}_{\text{ISP}}$ and $\mathcal{M}_{\text{CP}}$ decide the prices through a \mbox{non-cooperative} game, then, at the second stage, $\mathcal{M}_{\text{user}}$ make the best response traffic usage decision according to the prices.\\
\indent Initially, suppose $\mathcal{M}_{\text{ISP}}$ charges $\mathcal{M}_{\text{CP}}$ using a \mbox{bandwidth-based} price $p_b$ and charges $\mathcal{M}_{\text{user}}$ using a flat price $\tau$. We assume the equivalent \mbox{bandwidth-based} unit price of $\tau$ is the same as $p_b$. Thus, $\tau$ is often set based on a given $\xi_{\text{user}}$ ($\tau=\frac{v}{\xi_{\text{user}}}\cdot p_b$). Then, clearly, the profit of $\mathcal{M}_{\text{ISP}}$ is
\begin{equation}\label{isp-cs}
\begin{array}{rcl}
   \mathds{U}_{\text{ISP}}(p_b) & =& b_{\text{CP}}\cdot p_b+\tau-\mathbf{C}_{\text{ISP}}(v)\\
                           & =& \left(\frac{v}{\xi_{\text{CP}}}+\frac{v}{\xi_{\text{user}}}\right)\cdot p_b-\mathbf{C}_{\text{ISP}}(v),
\end{array}
\end{equation}
where $\mathbf{C}_{\text{ISP}}(\cdot)$ is a composite cost function \cite{ISP-cost}.\\
\indent For $\mathcal{M}_{\text{CP}}$, let $p_s$ be unit service price and $\mathbf{F}_{ad}(\cdot)$ be a \mbox{volume-based} advertisement fee function. Then, its profit is
\begin{equation}\label{cp-cs}
\begin{array}{rcl}
  \mathds{U}_{\text{CP}}(p_s) &=&  v\cdot p_s+\mathbf{F}_{ad}(v)-b_{\text{CP}}\cdot p_b-\mathbf{C}_{\text{CP}}(v) \\
                         &=&  v\cdot p_s+\mathbf{F}_{ad}(v)-\frac{v}{\xi_{\text{CP}}}\cdot p_b-\mathbf{C}_{\text{CP}}(v),
\end{array}
\end{equation}
where $\mathbf{C}_{\text{CP}}(\cdot)$ is a \mbox{volume-based} cost function.\\
\indent In addition, let $\mathbf{E}_{\text{user}}(v)$ be the experience value for $\mathcal{M}_{\text{user}}$ consuming content volume $v$.  Then, its utility is
\begin{equation}\label{user-cs}
\begin{array}{rcl}
  \mathds{U}_{\text{user}}(v) &=& \mathbf{E}_{\text{user}}(v)-(b_{\text{user}}\cdot p_b+v\cdot p_s)  \\
                           &=& \mathbf{E}_{\text{user}}(v)-(\frac{p_b}{\xi_{\text{user}}}+p_s)\cdot v.
\end{array}
\end{equation}
\indent In C/S network, a \mbox{three-player} game can characterize the interactions. $\mathcal{M}_{\text{ISP}}$ and $\mathcal{M}_{\text{CP}}$ act as leaders to price $\mathcal{M}_{\text{user}}$, which acts as follower to decide traffic usage.\\
\indent According to \emph{backward induction} for leader-follower game, we first analyze the second stage of this game, assuming that $\mathcal{M}_{\text{ISP}}$ and $\mathcal{M}_{\text{CP}}$ have set the prices at the first stage. \\
\textbf{The follower's problem}\\
\indent Given $p_b$ and $p_s$, $\mathcal{M}_{\text{user}}$ is going to maximize its utility in Eq.(\ref{user-cs}). By solving the follower's problem, we can obtain the service content consumed by $\mathcal{M}_{\text{user}}$ as
\begin{equation}\label{v-user}
    \hat{v}(p_b,p_s)=\min\{\underset{v}{\text{argmax}} \mathds{U}_{\text{user}} , b_{\text{user}}\cdot\xi_{\text{cp}}\}.
\end{equation}
which is the users' best response traffic usage decision within its purchased capacity. Let $o(v)=\frac{\text{d}\mathbf{E}_{\text{user}}(v)}{\text{d} v}$, $o(v)$ is a one-to-one mapping. Then based on our assumption that the network is underused, we have $\hat{v}(p_b,p_s)=o^{-1}(d\cdot p_b+p_s)$.\\
\textbf{The leaders' problems}\\
\indent Anticipating users will choose $v=\hat{v}(p_b,p_s)$ as its traffic usage, the leaders' problems become
  \begin{equation*}\label{v-n-isp}
    \text{For $\mathcal{M}_{\text{ISP}}$:~~~~}\underset{pb}{\text{max}} \mathds{U}_{\text{ISP}}(p_b,\hat{v}(p_b,p_s))
  \end{equation*} and
\begin{equation*}\label{v-n-cp}
    \text{For $\mathcal{M}_{\text{CP}}$:~~~~}\underset{ps}{\text{max}} \mathds{U}_{\text{CP}}(p_s,\hat{v}(p_b,p_s)).
\end{equation*}
Then a \mbox{two-player} \mbox{non-cooperative} game happens between $\mathcal{M}_{\text{ISP}}$ and $\mathcal{M}_{\text{CP}}$. They take turns to optimize their own objects $\mathds{U}_{\text{ISP}}$ and $\mathds{U}_{\text{CP}}$ by varying their decision variables $p_b$ and $p_s$, respectively, treating the other as constant.
\subsubsection{Game solution}
Let $(p_b^*,p_s^*)$ be the Nash Equilibrium, then, according to the definition of Nash Equilibrium, the solution turns out to be as follows
\begin{equation}\label{back-games}
\left\{
    \begin{array}{l}
      p_b^*= \underset{p_b}{\text{argmax}} \mathds{U}_{\text{ISP}}(p_b,\hat{v}(p_b,p_s^*)), \\
      p_s^*= \underset{p_s}{\text{argmax}} \mathds{U}_{\text{CP}}(p_b^*,\hat{v}(p_b^*,p_s)).
    \end{array}
\right.
\end{equation}
We have the following theorem on the simplified sufficient conditions of Nash Equilibrium for this problem.\\

\begin{theorem}\label{theorem1}
Let $(p_b^*,p_s^*)$ be the Nash Equilibrium defined in Eq.(\ref{back-games}) and $v^*=\hat{v}(p_b^*,p_s^*)$. Then, let
\begin{equation}\label{define}
\begin{array}{l}
\Phi_1(v)=c\cdot v \cdot \frac{1}{d}\cdot \frac{\text{d} o(v)}{\text{d} v}-\frac{\text{d} \mathbf{C}_{\text{ISP}}(v)}{\text{d} v},\\
\Phi_2(v)=v \cdot \frac{\text{d} o(v)}{\text{d} v}+ \frac{\text{d} \mathbf{F}_{ad}(v)}{\text{d} v}-\frac{\text{d} \mathbf{C}_{\text{CP}}(v)}{\text{d} v}.
 \end{array}
\end{equation}
It must satisfy the following two conditions

\begin{description}
  \item[(i).]  $o(v^*)+\Phi_1(v^*)+\Phi_2(v^*)=0,$\\
  \item[(ii).] $(\frac{c}{d}\cdot \frac{\text{d}o(v)}{\text{d} v}+\frac{\text{d}\Phi_1(v)}{\text{d} v})|_{v^*}<0$ ~~and\\
      $(\frac{\text{d} o(v)}{\text{d} v}+\frac{\text{d} \Phi_2(v)}{\text{d} v})|_{v^*}<0,$
\end{description}
\end{theorem}
where $c=\frac{1}{\xi_{\text{user}}}+\frac{1}{\xi_{\text{cp}}},$ and $e=\frac{1}{\xi_{\text{cp}}}$.
\begin{proof}
Obviously we have $o(v^*)=d\cdot p_b^*+p_s^*$ according to the first order condition of Eq.(\ref{user-cs}).\\
\indent According to the definition of Nash Equilibrium, $p_b^*$ should be the best response to $p_s^*$ and vice versa. Since we do not consider the cases where the maximum profit happens at the boundary, we must have
\begin{equation}\label{orign-con}
    \begin{array}{lllrrr}
      \frac{\partial\mathbf{E}_{\text{ISP}}(p_b,p_s^*)}{\partial p_b}|_{p_b^*}&=&0,~~~ \frac{\partial\mathbf{E}^2_{\text{ISP}}(p_b,p_s^*)}{\partial p_b}|_{p_b^*}&<&0,\\
      \frac{\partial\mathbf{E}_{\text{CP}}(p_b^*,p_s)}{\partial p_s}|_{p_s^*}&=&0,~~~
      \frac{\partial\mathbf{E}^2_{\text{CP}}(p_b^*,p_s)}{\partial p_s}|_{p_s^*}&<&0.
    \end{array}
\end{equation}
Moreover, because $v=o^{-1}(d\cdot p_b+p_s)$ exists, we have
\begin{equation}\label{new-con}
\nonumber
    \begin{array}{l}
      \frac{\partial\mathbf{E}_{\text{ISP}}(v,p_s^*)}{\partial v}|_{v^*}= \frac{\partial\mathbf{E}_{\text{ISP}}(p_b,p_s^*)}{\partial p_b}|_{p_b^*}\cdot\frac{\partial p_b}{\partial v}|_{v^*},\\
      \frac{\partial p_b}{\partial v}|_{v^*}=\frac{1}{d}\frac{\text{d} o(v)}{\text{d} v}|_{v^*}\neq 0,\\
        \frac{\partial\mathbf{E}_{\text{CP}}(v,p_b^*)}{\partial v}|_{v^*}= \frac{\partial\mathbf{E}_{\text{ISP}}(p_b^*,p_s)}{\partial p_s}|_{p_s^*}\cdot\frac{\partial p_s}{\partial v}|_{v^*},\\
      \frac{\partial p_s}{\partial v}|_{v^*}=\frac{\text{d} o(v)}{\text{d} v}|_{v^*}\neq 0.\\
    \end{array}
\end{equation}
We can apply the above properties on conditions in Eq.(\ref{orign-con}) and rewrite them as follows
\begin{equation}\label{renew-con}
    \begin{array}{lllrrr}
      \frac{\partial\mathbf{E}_{\text{ISP}}(v,p_s^*)}{\partial v}|_{v^*}&=&0,~~~ \frac{\partial\mathbf{E}^2_{\text{ISP}}(v,p_s^*)}{\partial v}|_{v^*}&<&0,\\
      \frac{\partial\mathbf{E}_{\text{CP}}(p_b^*,v)}{\partial v}|_{v^*}&=&0,~~~
      \frac{\partial\mathbf{E}^2_{\text{CP}}(p_b^*,v)}{\partial v}|_{v^*}&<&0.
    \end{array}
\end{equation}
From Eq.(\ref{renew-con}), Eq.(\ref{isp-cs}), Eq.(\ref{cp-cs}) and $o(v^*)=d\cdot p_b^*+p_s^*$, we have
\begin{equation}
\nonumber
    \begin{array}{l}
     c\cdot p_b^*+c\cdot v^* \cdot \frac{1}{d}\cdot \frac{\text{d} o(v)}{\text{d} v}|_{v^*}-\frac{\text{d} \mathbf{C}_{\text{ISP}}(v)}{\text{d} v}|_{v^*}=0,\\
     p_s^*+ v^*\cdot \frac{\text{d} o(v)}{\text{d} v}|_{v^*}+\frac{\text{d} \mathbf{F}_{ad}(v)}{\text{d} v}|_{v^*}-e\cdot p_b-\frac{\text{d} \mathbf{C}_{\text{CP}}(v)}{\text{d} v}|_{v^*}=0.
    \end{array}
\end{equation}
It can be further simplified based on Eq.(\ref{define}) as follows
\begin{equation}\label{simple}
    \begin{array}{l}
     c\cdot p_b^*+\Phi_1(v^*) =0,\\
     p_s^*+\Phi_2(v^*)-e\cdot p_b^* =0.
    \end{array}
\end{equation}
$v^*$ must satisfy the following condition in order to be the traffic usage at a Nash Equilibrium
\begin{equation}\label{ne-v}
o(v^*)+\Phi_1(v^*)+\Phi_2(v^*)=0.
\end{equation}
Second order conditions in Eq.(\ref{renew-con}) can be simplified as

\begin{equation}\label{simple-second}
    \begin{array}{l}
      (\frac{c}{d}\cdot \frac{\partial o(v)}{\partial (v)}+\frac{\partial \Phi_1(v^*)}{\partial v})|_{v^*}<0,~~\text{and}\\
      (\frac{\partial o(v)}{\partial (v)}+\frac{\partial \Phi_2(v^*)}{\partial v})|_{v^*}<0.
          \end{array}
\end{equation}
\end{proof}
This theorem presents a way to compute Nash Equilibrium which represents the steady state of this network market (denoted as State 0).
\subsubsection{Example and equilibrium analysis}
In a practical system, $\mathbf{C}_{\text{CP}}(\cdot)$ is often increasing and concave. $\mathbf{C}_{\text{ISP}}(\cdot)$ is often continuous increasing. When $v$ is small, the growth rate of this cost decreases with a larger $v$ and when $v$ is large, the growth rate increases with a even larger $v$ due to congestion (We use congestion cost to indicate potential expansion cost for ISPs, which increase fast when $v$ approaches to $b_{\text{ISP}}$ \cite{ISP-cost}). Generally, $\mathbf{F}_{ad}(\cdot)$ and $\mathbf{E}_{\text{user}}(\cdot)$ are increasing and concave.

\begin{figure}[htp]
\centering
\includegraphics[width=2.2in]{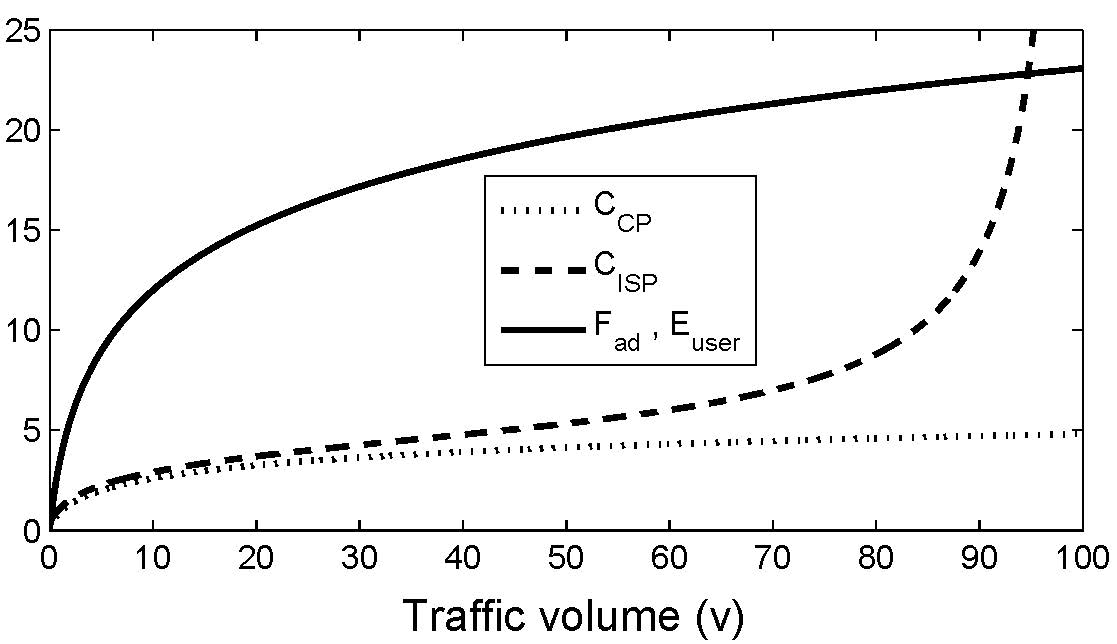}
\caption{Change trends of $\mathbf{C}_{\text{CP}}(v)$, $\mathbf{C}_{\text{ISP}}(v)$, $\mathbf{F}_{ad}(v)$ and $\mathbf{E}_{\text{user}}(v)$.}\label{cost-function}
\end{figure}

\indent Without loss of generality, we take $\mathbf{C}_{\text{CP}}(v)=\ln{(v+1)}+0.2$, $\mathbf{C}_{\text{ISP}}(v)=\ln{(v+1)}+100(\frac{1}{b_{\text{ISP}}-v}-\frac{1}{b_{\text{ISP}}})+0.4$, $\mathbf{F}_{ad}(v)=5\ln(v+1)$ and $\mathbf{E}_{\text{user}}(v)=5\ln(v+1)$, which have the above mentioned properties (See Fig.~\ref{cost-function}). Then, according to Eq.(\ref{v-user}) and our assumption, we have $\hat{v}(p_b,p_s)=\frac{5}{d\cdot p_b+p_s}-1$ which will not exceed the capacity. Based on \emph{Theorem}~\ref{theorem1}, we can directly derive the Nash Equilibrium in \mbox{closed-form}. Then, we can further study the sensitivity of $v^*$ on variables $\xi_{\text{CP}}$ and $\xi_{\text{user}}$. Here, we assume $b_{\text{ISP}}=100$, $0.1\leq\xi_{\text{user}},\xi_{\text{CP}}\leq0.75$ and $\xi_{\text{user}}\leq 0.4\cdot\xi_{\text{CP}}$, then we find that the smaller ratio between $\xi_{\text{user}}$ and $\xi_{\text{CP}}$, the higher $v^*$.\\
\indent Assume $\xi_{\text{CP}}=0.75$ and $\xi_{\text{user}}=0.25$ (similar to Norton's prediction on bandwidth usage ratios \cite{usage-ratio}). The corresponding utilities of participators are $(\mathds{U}_{\text{ISP}}, \mathds{U}_{\text{CP}}, \mathds{U}_{\text{user}})=(2.2438, 3.9942, 2.3281)$.
\subsection{P2P-involved Profit Computing Model}
\label{subsection-p2p}

One important work of this paper is to measure and quantify \mbox{P2P} traffic's impact on the network economic market under traditional pricing mechanism. It help us to analyze and predict potential changes of participators' decisions.\\
\indent Based on the results in last subsection, we first analyze the growing impact of P2P traffic on the profits or utilities of Internet participators when the pricing has not changed, i.e., \mbox{State 1}. $\mathcal{M}_{\text{ISP}}$ will bear an increasingly large burden with the growing \mbox{P2P} traffic based on Eq.(\ref{isp-cs}). Therefore we give an analysis of $\mathcal{M}_{\text{ISP}}$'s reactive behavior conditionally and study its corresponding aftermath, i.e., State 2. Finally, we give a state transformation graph to summarize these possible \mbox{non-cooperative} market states and their transition conditions.
\subsubsection{State 1}\label{section3-state1}
In \mbox{peer-assisted} network, we assume $v_{\text{cs}}$ will not be impacted by the emergence of P2P traffic (i.e. $v_{\text{cs}}=v_{\text{cs}}^*=v^*\cdot (1-\alpha)$). It is reasonable when people give priority to inelastic needs of traditional Internet services (such as email and web) which are unlikely to become \mbox{P2P-assisted}.\\
\indent Compared with C/S mode, \mbox{P2P} can improve the experience of $\mathcal{M}_{\text{user}}$ because of its scalability (especially as the accelerated service speed). So let $\widehat{\mathbf{E}}_{\text{user}}$ be $\mathcal{M}_{\text{user}}$'s new experience value for content downloading profile $v=v_{\text{p2p}}+v_{\text{cs}}^*$ and we assume $\widehat{\mathbf{E}}_{\text{user}}(v)>\mathbf{E}_{\text{user}}(v)$ as along as $v>v_{\text{cs}}^*$ (i.e., $v_\text{p2p}>0$). Let $a$ be the experience accelerate factor of P2P traffic (which is related with $\beta$ and always satisfies $a>1$), so we have $\widehat{\mathbf{E}}_{\text{user}}(v)=\mathbf{E}_{\text{user}}(a\cdot(v-v_{\text{cs}}^*)+v_{\text{cs}}^*)$. Indeed, we simply assume that $a$ and $\beta$ satisfy a linear relationship. So we create fitting curve for $a$ based on two empirical points $(\beta,a)=(0.3,4)$ (i.e., when 70\% \mbox{PCP} content is provided by \mbox{P2P}, users' experience will expand to 4 times compared with under \mbox{C/S} mode.) and $(\beta,a)=(1,1)$ (i.e., when all the \mbox{PCP} content is provided by servers, the calculation of such experience is the same as under \mbox{C/S} mode). Then, we get $a=1+\frac{30}{7}(1-\beta)$.

\begin{remark}
Intuitively, $1-\beta$ reflects \mbox{P2P}'s power and when it becomes larger, the performance of \mbox{P2P} service will become better as its distributed sharing nature. So we assume $k$ increases with $1-\beta$. As \mbox{PCPs}' servers guarantee system stability, they are generally indispensable (i.e., $\beta>0$).
\end{remark}

\indent Often, $\mathcal{M}_{\text{ISP}}$ charges $\mathcal{M}_{\text{user}}$ at a flat price. Suppose new average bandwidth usage ratio $\xi_{\text{user}}^*$ cannot exceed $\xi_{\text{CP}}$ as discussed in Section \ref{network-model}. Let $\widetilde{v}_{\text{p2p}} =\frac{b_{\text{user}}^*\cdot\xi_{\text{CP}}-v_{\text{cs}}^*}{2-\beta}$. As long as $v_{\text{p2p}}\leq\widetilde{v}_{\text{p2p}} $, the fee charged from $\mathcal{M}_{\text{user}}$ will be kept at $\tau=b_{\text{user}}^*\cdot p_b^*$; when $v_{\text{p2p}}>\widetilde{v}_{\text{p2p}} $, we assume $\mathcal{M}_{\text{ISP}}$ will charge on the excessive volume $(v_{\text{p2p}}-\widetilde{v}_{\text{p2p}})\cdot(2-\beta)$ based on a \mbox{volume-based} pricing. For \mbox{bandwidth-based} price $p_b^*$, its equivalent \mbox{volume-based} price is $\frac{p_b^*}{\xi_{\text{user}}}$. Thus, the utility of $\mathcal{M}_{\text{user}}$ becomes
\begin{equation}\label{user-p2p}
\mathds{U}_{\text{user}} =
\left\{
  \begin{array}{l}
    \widehat{\mathbf{E}}_{\text{user}}(v)-v\cdot p_s^*-\tau, ~\hbox{ if $v_{\text{p2p}}\leq\widetilde{v}_{\text{p2p}}$;} \\[0.2cm]
    \widehat{\mathbf{E}}_{\text{user}}(v)-v\cdot p_s^*-\tau-\\
     \ \ (v_{\text{p2p}}-\widetilde{v}_{\text{p2p}})\cdot(2-\beta)\cdot \frac{p_b^*}{\xi_{\text{user}}}, ~ \hbox{otherwise.}
  \end{array}
\right.
\end{equation}
Here, $\mathcal{M}_{\text{user}}$ will decide $v_{\text{p2p}}^{\text{S1}}$ (since $v=v_{\text{p2p}}+v_{\text{cs}}^*$ based on our assumption) to maximize $\mathds{U}_{\text{user}}$.Then, based on $v_{\text{p2p}}^{\text{S1}}$, we can get $\mathds{U}_{\text{CP}}$ and $\mathds{U}_{\text{ISP}}$ as follows.\\
\indent For $\mathcal{M}_{\text{CP}}$, $\mathds{U}_{\text{CP}}$ will become
\begin{equation}\label{u-cp-p2p}
      \begin{array}{l}
         \mathds{U}_{\text{CP}} = v^{\text{S1}}\cdot p_s^*+\mathbf{F}_{ad}(v^{\text{S1}})-\frac{v_{\text{p2p}}^{\text{S1}}\cdot \beta+v_{\text{cs}}^*}{\xi_{\text{CP}}}\cdot p_b^*-\widehat{\mathbf{C}}(v^{\text{S1}}),
      \end{array}
\end{equation}
where $v^{\text{S1}}=v_{\text{p2p}}^{\text{S1}}+v_{\text{cs}}^*$ and $\frac{v_{\text{p2p}}^{\text{S1}}\cdot \beta+v_{\text{cs}}^*}{\xi_{\text{CP}}}$ denotes the bandwidth purchased by  $\mathcal{M}_{\text{CP}}$ from $\mathcal{M}_{\text{ISP}}$ when the $\beta$ proportion traffic is provided by their own servers. Similar to $\widehat{\mathbf{E}}_{\text{user}}(v)$, here we define $\widehat{\mathbf{C}}_{\text{CP}}(v)=\mathbf{C}_{\text{CP}}((v-v_{\text{cs}}^*)\cdot \beta+v_{\text{cs}}^*)$ ($0<\beta\leq1$) to measure the cost alleviated through \mbox{P2P-assisting}.\\
\indent Accordingly, $\mathds{U}_{\text{ISP}}$ will become
\begin{equation}\label{u-isp-p2p}
\mathds{U}_{\text{ISP}} =
\left\{
  \begin{array}{l}
    \tau+\frac{v_{\text{p2p}}^{\text{S1}}\cdot \beta+v_{\text{cs}}^*}{\xi_{\text{CP}}}\cdot p_b^*-\mathbf{C}_{\text{ISP}}(v^{\text{S1}}), \hbox{ ~if $v_{\text{p2p}}^{\text{S1}}<=\widetilde{v}_{\text{p2p}}$;} \\
   \tau+(v_{\text{p2p}}^{\text{S1}}-\widetilde{v}_{\text{p2p}})\cdot(2-\beta)\cdot \frac{p_b^*}{\xi_{\text{user}}}+\\
    \ \ \frac{v_{\text{p2p}}^{\text{S1}}\cdot \beta+v_{\text{cs}}^*}{\xi_{\text{CP}}}\cdot p_b^*-\mathbf{C}_{\text{ISP}}(v^{\text{S1}}), ~~~~~\hbox{otherwise.}
  \end{array}
\right.
\end{equation}
\indent Following the foregoing example, we assume $\alpha=0.6$ ($\beta=0.3$ as previously assumed). Then by solving the \mbox{user-side} optimization problem, we can deduce the optimal point $v_{\text{p2p}}^{\text{S1}}$. It is clear that $v_{\text{p2p}}^{\text{S1}}> v^*\cdot\alpha$. In addition, we can see that $v_{\text{p2p}}^{\text{S1}}$ increases to but does not exceed $\widetilde{v}_{\text{p2p}}$. This implies that $\mathcal{M}_{\text{user}}$ will try to use up its original bandwidth bought from $\mathcal{M}_{\text{ISP}}$ with a flat price but not to buy additional bandwidth. Then, we can obtain $(\mathds{U}_{\text{ISP}}, \mathds{U}_{\text{CP}}, \mathds{U}_{\text{user}})=(1.5964, 7.2021, 9.6230)$ for State 1. Compared with \mbox{State 0}, $\mathds{U}_{\text{CP}}$ increases by $80.31\%$, while $\mathds{U}_{\text{ISP}}$ decreases by $28.85\%$. Thus, motivated by profit increase, some CPs will adopt P2P technology and become PCPs. Then, the overall system will change from \mbox{State 0} to \mbox{State 1}.

\begin{remark}
Economically, the only condition for system to change from \mbox{State 0} to \mbox{State 1} is that under traditional pricing mechanism, $\mathds{U}_{\text{CP}}^{\text{S1}}>\mathds{U}_{\text{CP}}^{\text{S0}}$. According to Eqs. (\ref{user-cs})~and~(\ref{user-p2p}), it is easily proved that $v_{\text{p2p}}^{\text{S1}}+v_{\text{cs}}*> v^*$ is always true.
\end{remark}

\subsubsection{State 2}
For $\mathcal{M}_{\text{ISP}}$, one main reason for its decreasing profit is that it charges $\mathcal{M}_{\text{user}}$ at a flat price, which leads to P2P \mbox{free-riding}. To defeat such \mbox{free-riders}, one effective way is to change the original flat pricing model to a \mbox{volume-based} pricing model  \cite{P2Pfea,uplink-pricing,survey}. Like in \mbox{State 2}, we adopt $\frac{p_b^*}{\xi_{\text{user}}}$ as the \mbox{volume-based} price. Then, the utility of $\mathcal{M}_{\text{user}}$ becomes
\begin{equation}\label{u-user-usage}
\begin{array}{l}
\mathds{U}_{\text{user}}=\widehat{\mathbf{E}}_{\text{user}}(v)-v\cdot p_s^*-
\left[v_{\text{p2p}}\cdot(2-\beta)+v_{\text{cs}}^*\right]\cdot{\frac{p_b^*}{\xi_{\text{user}}}}.
\end{array}
\end{equation}
$\mathcal{M}_{\text{user}}$ chooses $v_{\text{p2p}}^{\text{S2}}$ to obtain the feasible optimal traffic usage. Then, the utilities of $\mathcal{M}_{\text{ISP}}$ and $\mathcal{M}_{\text{CP}}$ can thus be solved. For $\mathds{U}_{\text{CP}}$, the computation method is equal to Eq.(\ref{u-cp-p2p}). Accordingly, $\mathds{U}_{\text{ISP}}$ becomes
\begin{equation}\label{u-isp-usage}
\begin{array}{rcl}
\mathds{U}_{\text{ISP}}&=&\left[v_{\text{p2p}}^{\text{S2}}\cdot(2-\beta)+v_{\text{cs}}^*\right]\cdot{\frac{p_b^*}{\xi_{\text{user}}}} +\frac{v_{\text{p2p}}^{\text{S2}}\cdot\beta+v_{\text{cs}}^*}{\xi_{\text{CP}}}\cdot p_b^*- \\
 & &\mathbf{C}_{\text{ISP}}(v^{\text{S2}}),
\end{array}
\end{equation}
where $v^{\text{S2}}=v_{\text{p2p}}^{\text{S2}}+v_{\text{cs}}^*$.\\
\indent Correspondingly, we have $(\mathds{U}_{\text{ISP}}, \mathds{U}_{\text{CP}}, \mathds{U}_{\text{user}})=(3.5180, 5.6450, 5.1712)$. Therefore, after $\mathcal{M}_{\text{ISP}}$ adopts a \mbox{volume-based} pricing model, $\mathds{U}_{\text{ISP}}$ increases by $120.66\%$, while $\mathds{U}_{\text{CP}}$ decreases by $29.42\%$. Thus, motivated by profit increase, $\mathcal{M}_{\text{ISP}}$ will change its flat pricing model on $\mathcal{M}_{\text{user}}$ to a \mbox{volume-based} one. Then, the overall system will change from \mbox{State 1} to \mbox{State 2}. Since $\mathds{U}_{\text{CP}}^{\text{S2}}>\mathds{U}_{\text{CP}}^{\text{S0}}$, $\mathcal{M}_{\text{PCP}}$ still benefits from P2P technology and will not take further actions against $\mathcal{M}_{\text{ISP}}$ except for a better choice.
\begin{remark}
The two conditions for the overall system to change from State 1 to State 2 are $\mathds{U}_{\text{ISP}}^{\text{S1}}<\mathds{U}_{\text{ISP}}^{\text{S0}}$ and $\mathds{U}_{\text{ISP}}^{\text{S2}}>\mathds{U}_{\text{ISP}}^{\text{S1}}$, respectively. If $\mathds{U}_{\text{ISP}}^{\text{S1}}>\mathds{U}_{\text{ISP}}^{\text{S0}}$, $\mathcal{M}_{\text{ISP}}$ will benefit from P2P technology. However, according to Eqs.(\ref{user-p2p})~and~(\ref{u-user-usage}), it is easy to prove that $v_{\text{p2p}}^{\text{S2}}<v_{\text{p2p}}^{\text{S1}}$ is always true. Then, $\mathcal{M}_{\text{ISP}}$ does not need to change its pricing model on $\mathcal{M}_{\text{user}}$.
\end{remark}
\begin{remark}
For $\mathcal{M}_{\text{PCP}}$, if $\mathds{U}_{\text{CP}}^{\text{S2}}<\mathds{U}_{\text{CP}}^{\text{S0}}$ (Since traffic demand is suppressed under $\mathcal{M}_{\text{ISP}}$'s \mbox{user-side} new pricing mode, the saved cost cannot offset the reduced income), it may give up P2P technology as the reduced profit. Then, the overall system will be forced to change from \mbox{State 2} to \mbox{State 0}.
\end{remark}

\subsubsection{Discussion and Non-cooperative State Analysis}
\label{state-transmission-analysis}

\indent In this subsection, we describe another two possible \mbox{non-cooperative} states. These help us to analyze \emph{how P2P technology will affect the network participators' behaviors and utilities.} Through analyzing these states, we can quantify the profits and predict the possible profit changing trends of network participators under fixed traffic profile (i.e., $\alpha$ and $\beta$) and unchanged pricing levels (i.e., $p_b$ and $p_s$). The reasons are as follows:
\begin{description}
  \item[a)] We need to study how \mbox{P2P} traffic impacts profit distribution among these players if \mbox{non-p2p} traffic (i.e., traffic of $\mathcal{M}^r_{\text{CP}}$) is treated the same as at \mbox{State 0} by $\mathcal{M}_{\text{ISP}}$ and $\mathcal{M}_{\text{user}}$;
  \item[b)] $\mathcal{M}^r_{\text{CP}}$ decides $\beta$ mostly based on its own technology and network situation while minimizing its cost, rather than based on complex economic computation.
\end{description}

\begin{figure}[htp]
\centering
\includegraphics[width=1.4in]{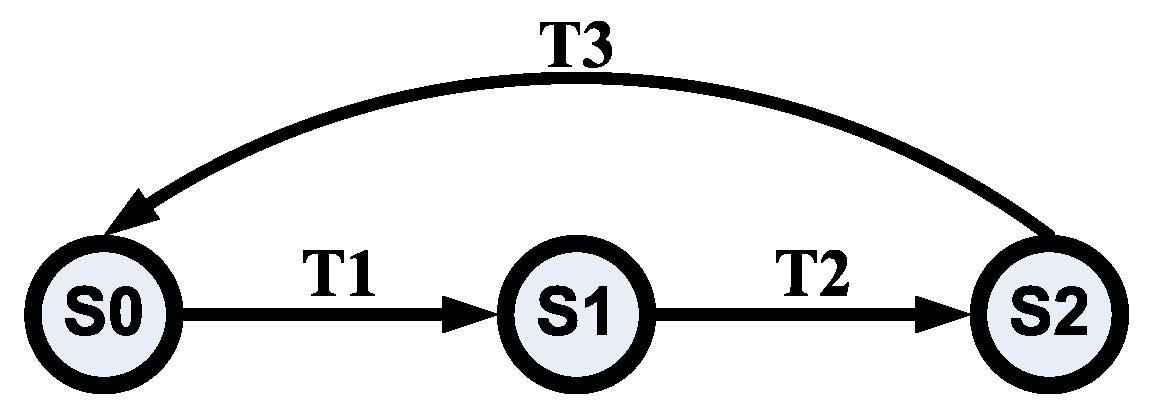}
\caption{State transformation among States 0, 1 and 2 (We use S0, S1 and S2 for short). The conditions for the three transitions T1, T2 and T3 are: (1) T1: $\mathds{U}_{\text{CP}}^{\text{S1}}>\mathds{U}_{\text{CP}}^{\text{S0}}$; (2) T2: $\mathds{U}_{\text{ISP}}^{\text{S1}}<\mathds{U}_{\text{ISP}}^{\text{S0}}$ and $\mathds{U}_{\text{ISP}}^{\text{S2}}>\mathds{U}_{\text{ISP}}^{\text{S1}}$; (3) T3: $\mathds{U}_{\text{CP}}^{\text{S2}}<\mathds{U}_{\text{CP}}^{\text{S0}}$.}\label{state-transimition}
\end{figure}

\indent Fig.~\ref{state-transimition} summarizes the state transformation conditions among States 0, 1 and 2. We summarize all possible equilibrium states (i.e., subgame perfect Nash equilibriums, SPNEs) that system can attain and the conditions for each one.

\begin{figure}[htp]
\centering
\includegraphics[width=2.2in]{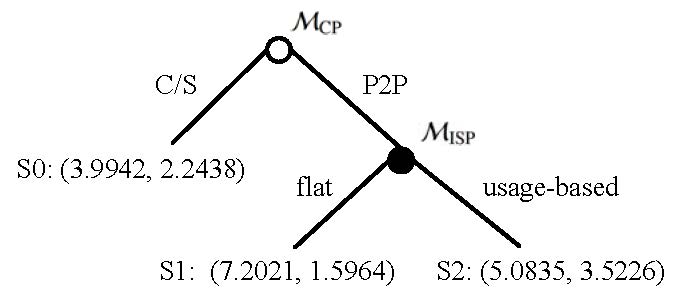}
\caption{An example of this dynamic game, on the leaf nodes are the utility values of ($\mathds{U}_{\text{CP}},\mathds{U}_{\text{ISP}}$) at different states.}\label{game-tree1}
\end{figure}

\indent For the forgoing example, its \emph{game tree} is illustrated in Fig.~\ref{game-tree1}. As this tree shows, the game starts from an empty circle and $\mathcal{M}_{\text{CP}}$ decides whether to adopt \mbox{P2P} technology. If so, the game goes to the filled circle. Then, $\mathcal{M}_{\text{ISP}}$ decides which pricing model will be used to charge $\mathcal{M}_{\text{user}}$, i.e., ``flat'' or ``usage-based''. Afterwards, the game is over. Based on \emph{backward induction}, we get \mbox{(P2P, usage-based pricing)} as the \mbox{SPNE} and the equilibrium payoff vector is $(5.0835, 3.5226)$. We can verify that it satisfies the conditions for \mbox{State 2} to be the final state (i.e., T1 instead of T2 in Fig.~\ref{state-transimition}).

\begin{figure}[htp]
\centering
\subfigure[]{
\includegraphics[width=2.2in]{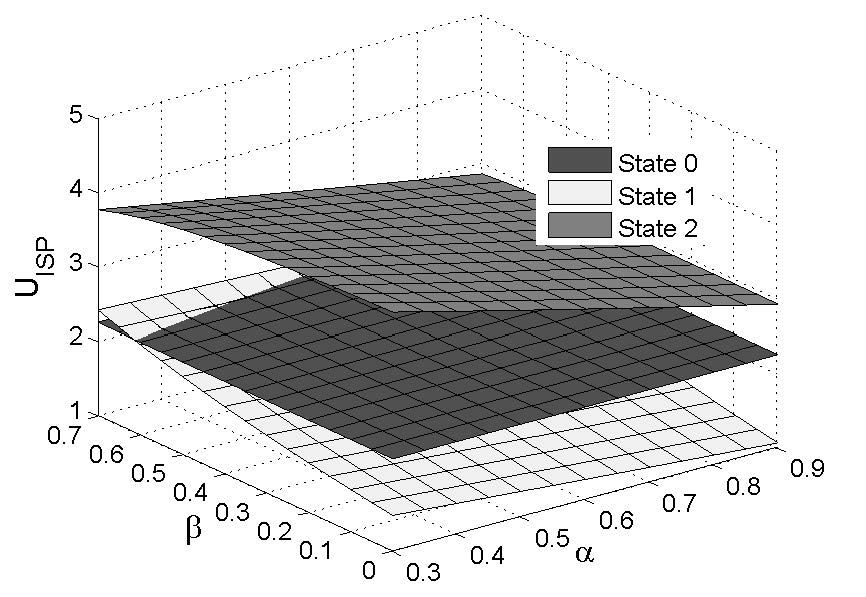}
\label{profit-ISP}
}
\subfigure[]{
\includegraphics[width=2.2in]{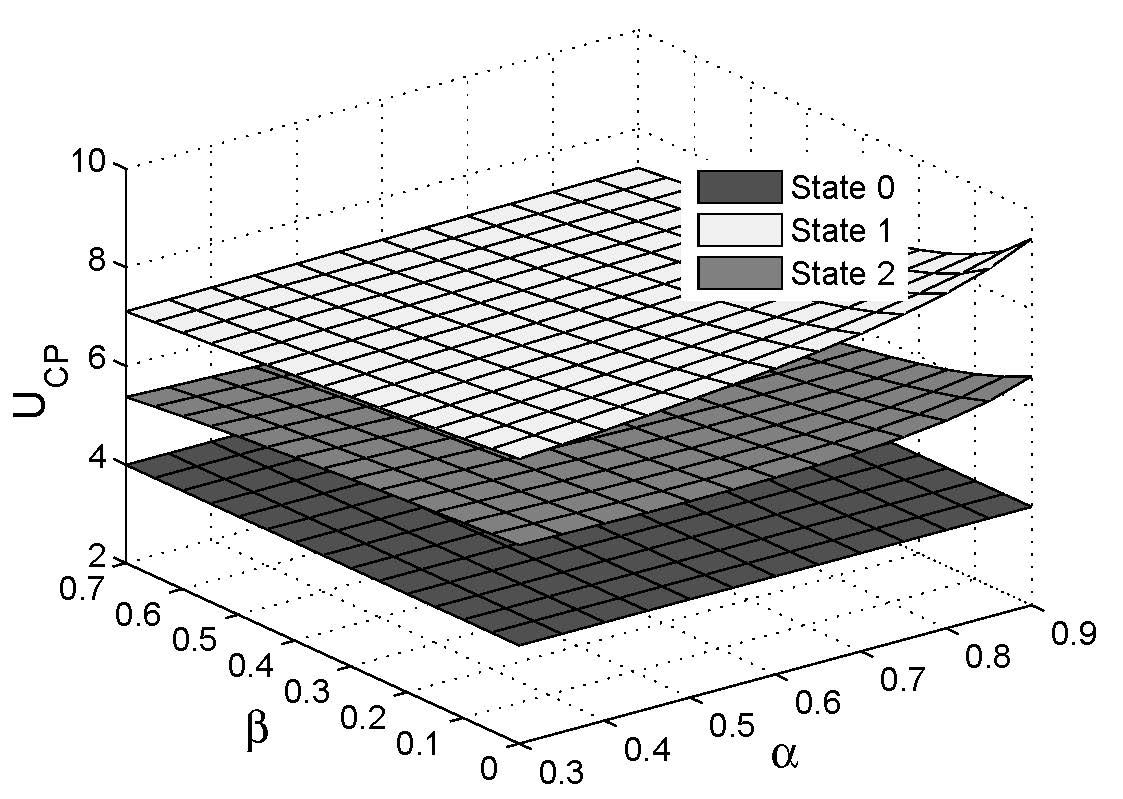}
\label{profit-CP}
}
\caption{(a) $\mathds{U}_{\text{ISP}}$ and (b) $\mathds{U}_{\text{CP}}$ for different $\alpha$ and $\beta$.}\label{profit}
\end{figure}

\indent In a practical system, using the state transformation conditions in Fig.~\ref{state-transimition}, we can conclude the conditions for each \mbox{SPNE}. Under certain condition, each state can be a proper Nash equilibrium.\\
\indent For different traffic profiles $(\alpha, \beta)$, we can correspondingly derive the utilities of $\mathcal{M}_{\text{ISP}}$ and $\mathcal{M}_{\text{CP}}$ as Fig.~\ref{profit} shows.
We plotted the initial equilibrium utilities computed in Section~\ref{basic-state0} as a comparison.\\
\indent Practically, $\beta$ is often smaller than 0.5. Then in Fig.~\ref{profit}, we can see the overall system finally stop at \mbox{State 2}, where $\mathcal{M}_{\text{ISP}}$ charges $\mathcal{M}_{\text{user}}$ using an \mbox{usage-based} pricing model. Here, $\mathds{U}_{\text{ISP}}^{\text{S2}}$ increases by $120.66\%$ compared with $\mathds{U}_{\text{ISP}}^{\text{S1}}$ and increases by $56.99\%$ compared with $\mathds{U}_{\text{ISP}}^{\text{S0}}$; $\mathds{U}_{\text{CP}}^{\text{S2}}$ reduces by $29.42\%$ compared with $\mathds{U}_{\text{CP}}^{\text{S1}}$ though increases by $27.27\%$ compared with  $\mathds{U}_{\text{CP}}^{\text{S0}}$.

\section{Cooperative Profit-distribution Model}
\label{newnbs}

\begin{figure}[htp]
\centering
\includegraphics[width=2.2in]{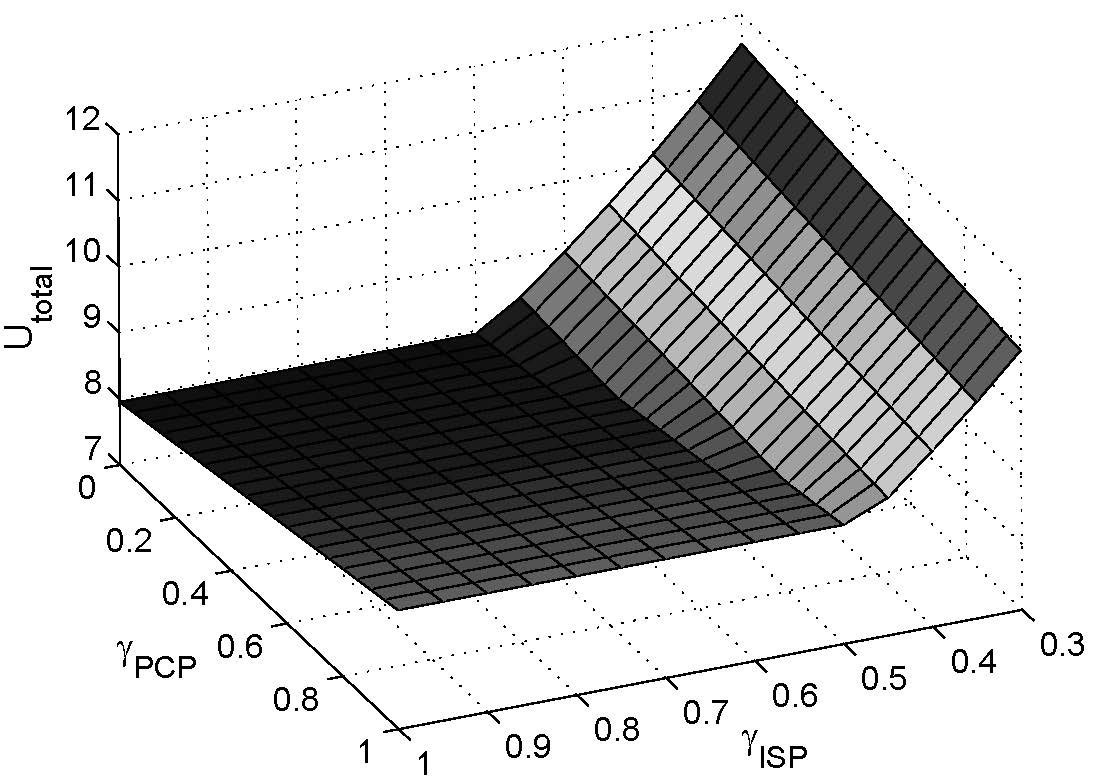}
\label{utotal}
\caption{$\mathds{U}_{\text{total}}$ for different $\gamma_{\text{ISP}}$ and $\gamma_{\text{PCP}}$ in the \mbox{peer-assisted} network with traffic profiles $(\alpha,\beta)=(0.6,0.3)$.}\label{cooper}
\end{figure}

In this section, we propose a cooperative profit distribution model based on the concept of Nash Bargaining Solution (NBS) \cite{NBS}, in which ISPs and PCPs cooperatively maximize their total profit and then fairly distribute profit based on NBS.\\
\indent Based on our analysis in the Section~\ref{section3-state1}, we notice that in the \mbox{peer-assisted} network, $\mathcal{M}_{\text{user}}$ may use up its original bandwidth bought from $\mathcal{M}_{\text{ISP}}$ at a flat price without buying additional bandwidth at a \mbox{volume-based} price. Here we consider the following cooperation: PCP coalition sells content at a discount rate $\gamma_{\text{PCP}}$ and ISP coalition charges the increased bandwidth bought by $\mathcal{M}_{\text{user}}$ at a discount rate $\gamma_{\text{ISP}}$. Both of them try to encourage $\mathcal{M}_{\text{user}}$ to consume more content and buy more bandwidth for P2P applications. Fig.~\ref{cooper} shows that ISPs and PCPs must cooperate to encourage $\mathcal{M}_{\text{user}}$ to consume more P2P content and to gain increased total profit.

\indent In this cooperation, $\mathds{U}_{\text{user}}$, $\mathds{U}_{\text{ISP}}$ and $\mathds{U}_{\text{CP}}$ become
\begin{equation*}\label{user-co}
\mathds{U}_{\text{user}} =
\left\{
  \begin{array}{l}
    \widehat{\mathbf{E}}_{\text{user}}(v)-(v_{\text{p2p}}\cdot\gamma_{\text{PCP}}+v_{\text{cs}}^*)\cdot p_s^*-\tau,\\
    \ \ \ \ \ \ \ \ \ \ \ \ \ \ \ \ \ \ \ \ \ \  \ \ \ \ \ \ \ \ \ \ \ \ \ \hbox{ if $v_{\text{p2p}}\leq\widetilde{v}_{\text{p2p}}$;} \\[0.2cm]
    \widehat{\mathbf{E}}_{\text{user}}(v)-(v_{\text{p2p}}\cdot\gamma_{\text{PCP}} +v_{\text{cs}}^*)\cdot p_s^*-\tau-\\ \ \ (v_{\text{p2p}}-\widetilde{v}_{\text{p2p}})\cdot(2-\beta)\cdot \frac{p_b^*}{\xi_{\text{user}}}\cdot\gamma_{\text{ISP}}, \ \ \hbox{otherwise.}
  \end{array}
\right.
\end{equation*}

\begin{equation*}\label{u-isp-co}
\mathds{U}_{\text{ISP}} =
\left\{
  \begin{array}{l}
    \tau+\frac{v_{\text{p2p}}\cdot \beta+v_{\text{cs}}^*}{\xi_{\text{CP}}}\cdot p_b^*-\mathbf{C}_{\text{ISP}}(v), \\
    \ \ \ \ \ \ \ \ \ \ \ \ \ \ \ \ \ \ \  \ \ \ \ \ \ \ \ \ \ \ \ \ \hbox{ if $v_{\text{p2p}}\leq\widetilde{v}_{\text{p2p}}$;} \\
   \tau+(v_{\text{p2p}}-\widetilde{v}_{\text{p2p}})\cdot(2-\beta)\cdot \frac{p_b^*}{\xi_{\text{user}}}\cdot\gamma_{\text{ISP}} +\\
    \ \ \ \ \ \frac{v_{\text{p2p}}\cdot \beta+v_{\text{cs}}^*}{\xi_{\text{CP}}}\cdot p_b^*-\mathbf{C}_{\text{ISP}}(v),\ \ \ \hbox{ otherwise.}
  \end{array}
\right.
\end{equation*}

\begin{equation*}\label{u-cp-co}
      \begin{array}{rcl}
         \mathds{U}_{\text{CP}}& = &(v_{\text{p2p}}\cdot\gamma_{\text{PCP}}+v_{\text{cs}}^*)\cdot p_s^*+\mathbf{F}_{ad}(v)-\\
            & & \frac{v_{\text{p2p}}\cdot \beta+v_{\text{cs}}^*}{\xi_{\text{CP}}}\cdot p_b^*-\widehat{\mathbf{C}}_{\text{CP}}(v).
      \end{array}
\end{equation*}
Here, a \mbox{leader-follower} game happens between the cooperative group and $\mathcal{M}_{\text{user}}$. The former changes $\gamma_{\text{ISP}}$ and $\gamma_{\text{PCP}}$ to maximize its total profit:
\begin{equation}\label{u-total}
\nonumber
\mathds{U}_{\text{total}} =\mathds{U}_{\text{ISP}}+\mathds{U}_{\text{CP}}.
\end{equation}
While $\mathcal{M}_{\text{user}}$ as the price taker changes $v_{\text{p2p}}$ to maximize $\mathds{U}_{\text{user}}$:
\begin{equation}\label{game-ispcp-user}
    \begin{array}{l}
      \text{for the~} \mathcal{M}_{\text{user}}: \\
      \ \ \ \ \ \ \ \ \ \widehat{v}=\underset{v_{\text{p2p}}}{\text{argmax}} {\mathds{U}_{\text{user}}}(\gamma_{\text{ISP}},\gamma_{\text{PCP}});\\
       \text{for the cooperative group}:\\
       \ \ \ \ \ \underset{\gamma_{\text{ISP}},\gamma_{\text{PCP}}}{\text{max}} {\mathds{U}_{\text{total}}}(\gamma_{\text{ISP}},\gamma_{\text{PCP}},\widehat{v}(\gamma_{\text{ISP}},\gamma_{\text{PCP}})).
    \end{array}
\end{equation}

\begin{figure}[htp]
\centering
\subfigure[]{
\includegraphics[width=2.2in]{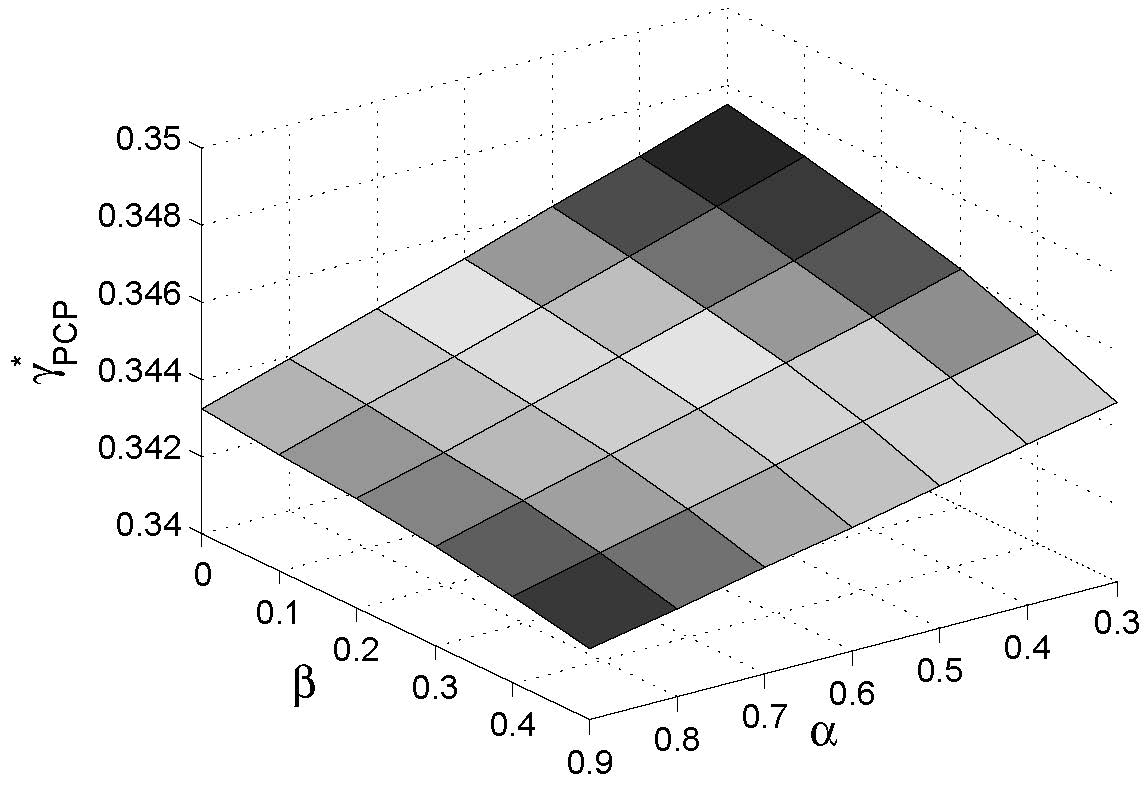}
\label{gamma-for-ab}
}
\subfigure[]{
\includegraphics[width=2.2in]{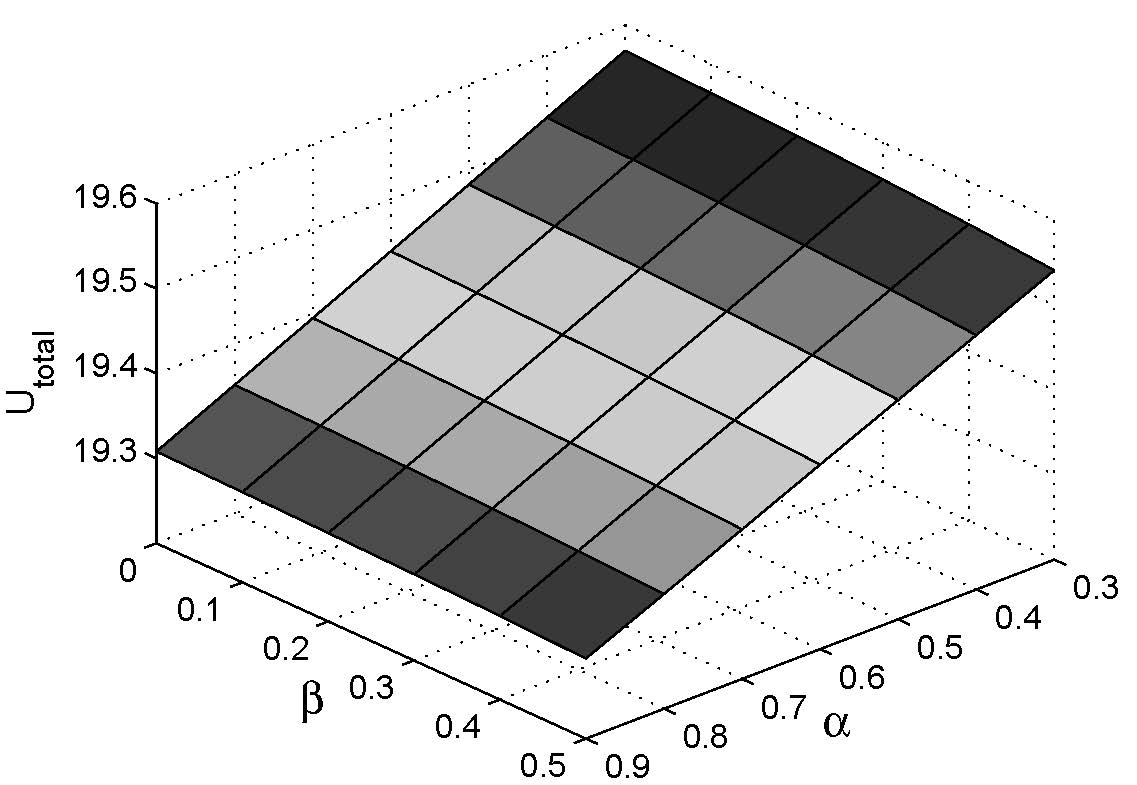}
\label{utotal}
}
\caption{Optimal $\gamma_{\text{PCP}}$ and $\mathds{U}_{\text{total}}$ for different $\alpha$ and $\beta$.}\label{for-ab}
\end{figure}

\indent By solving the above problem under different traffic profiles, we obtain the optimal values of $v_{\text{p2p}}$, $\gamma_{\text{PCP}}$ and $\mathds{U}_{\text{total}}$ illustrated in Fig.~\ref{for-ab}, which shows that $\gamma_{\text{PCP}}$ and $\mathds{U}_{\text{total}}$ decrease with the increase of $\alpha$ or $\beta$.

\begin{figure}[htp]
\centering
\subfigure[]{
\includegraphics[width=2.2in]{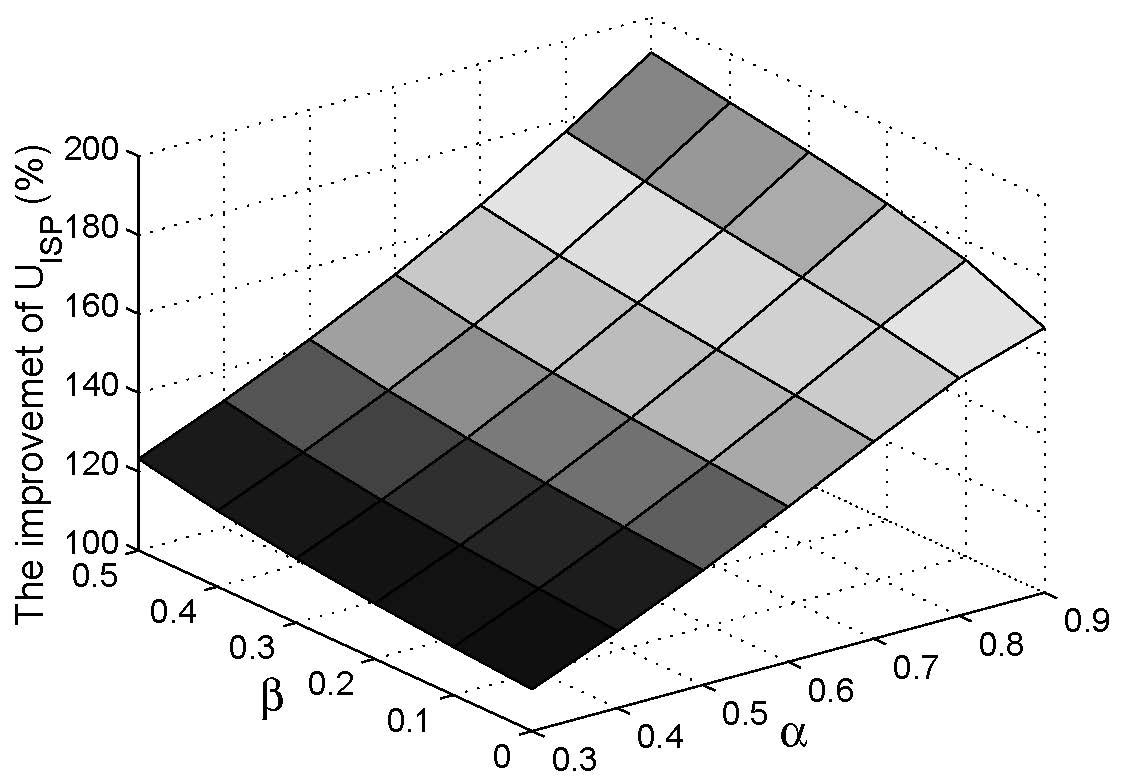}
\label{utotal-for-ab}
}
\subfigure[]{
\includegraphics[width=2.2in]{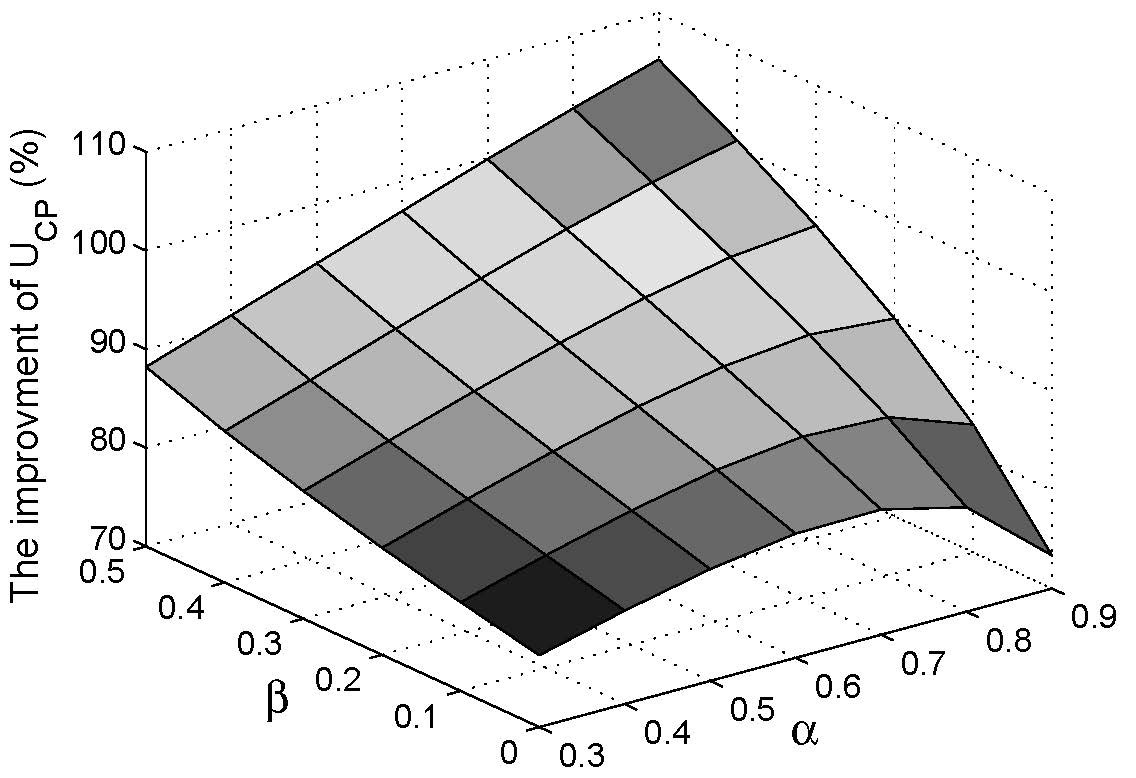}
\label{vp2p-for-ab}
}
\caption{The improvement of $\mathds{U}^{\text{S3}}_{\text{ISP}}$ and $\mathds{U}^{\text{S3}}_{\text{CP}}$ compared with the starting point for different $\alpha$ and $\beta$.}\label{improve-for-ab}
\end{figure}

\begin{figure}[!htp]
\centering
\includegraphics[width=2.2in]{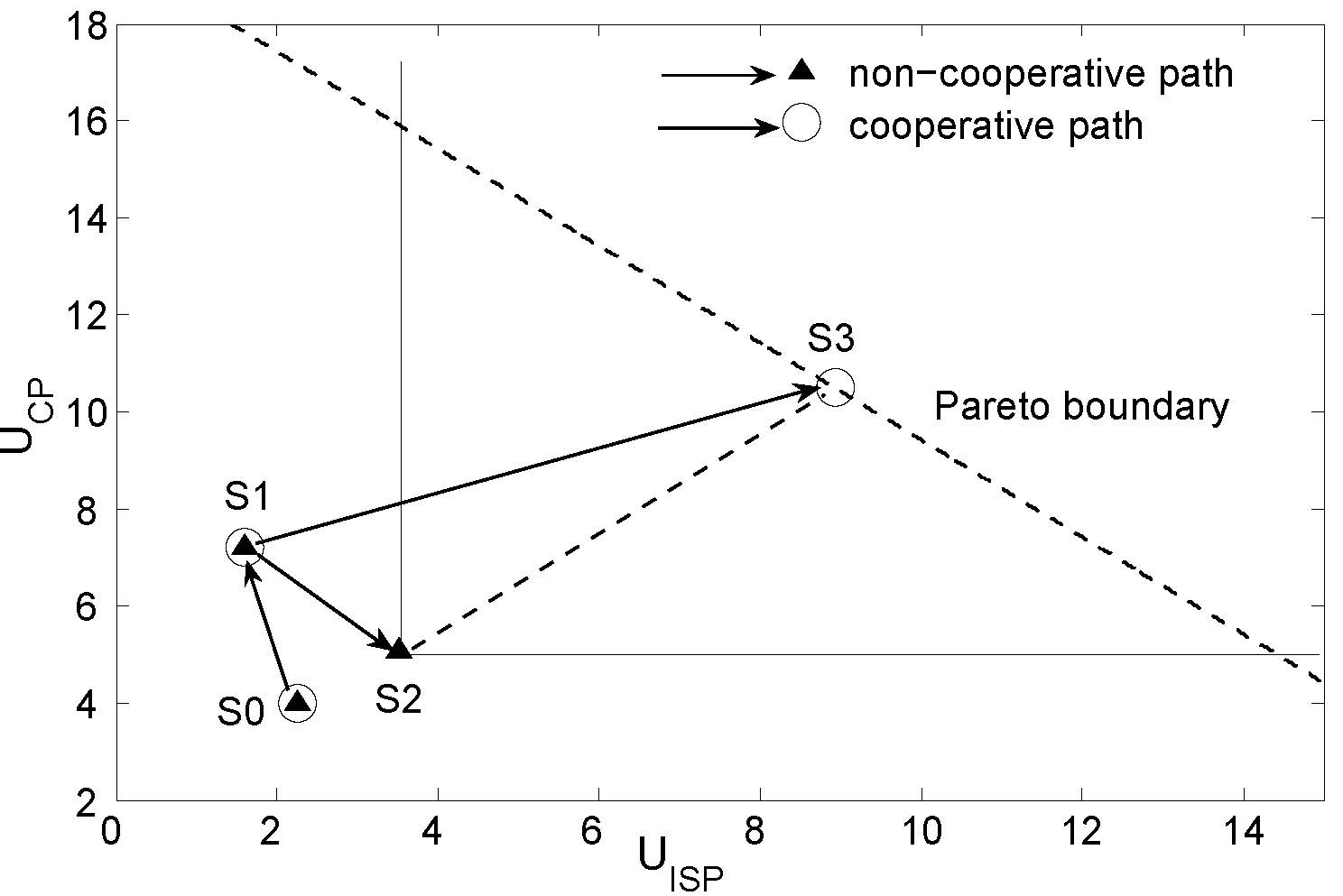}
\caption{An example of Nash bargaining between ISP and PCP coalitions $(\alpha,\beta)=(0.6,0.3)$.}\label{example}
\end{figure}

Then, for traffic profile $(\alpha,\beta)=(0.6,0.6)$, we can obtain the unique Stacklberg Equilibrium point where $\gamma_{\text{ISP}}^*=0$, $\gamma_{\text{PCP}}^*=0.3443$ and $v_{\text{p2p}}^{\text{S3}}=56.0140$. The results indicate that $\mathcal{M}_{\text{ISP}}$ will freely upgrade $\mathcal{M}_{\text{user}}$'s access bandwidth. Correspondingly, $(\mathds{U}_{\text{total}}^{\text{S3}},\mathds{U}_{\text{user}}^{\text{S3}})=(19.4287,19.0598)$. We can see that after cooperation, both $\mathds{U}_{\text{total}}$ and $\mathds{U}_{\text{user}}$ increase dramatically. Before PCP shares profit with ISP, $(\mathds{U}_{\text{ISP}}^{\text{S3'}},\mathds{U}_{\text{CP}}^{\text{S3'}})=(4.90593,14.5227)$. For all cases, $\mathds{U}_{\text{ISP}}+\mathds{U}_{\text{CP}}\leq\mathds{U}_{\text{total}}^{\text{S3}}$. Thus,
\begin{equation}
\mathds{U}_{\text{ISP}}+\mathds{U}_{\text{CP}}=\mathds{U}_{\text{total}}^{\text{S3}}
\end{equation}
is the corresponding Pareto boundary.\\
\indent Now, we are faced with an important question: \emph{How can ISP and PCP coalitions choose a fair point on the Pareto boundary as their profit distribution?} As discussed previously, without cooperation, their profit distribution may reach one of the following points (see Fig.~\ref{state-transimition}): $(\mathds{U}_{\text{ISP}}^{\text{S0}},\mathds{U}_{\text{CP}}^{\text{S0}})$, $(\mathds{U}_{\text{ISP}}^{\text{S1}},\mathds{U}_{\text{CP}}^{\text{S1}})$, or $(\mathds{U}_{\text{ISP}}^{\text{S2}},\mathds{U}_{\text{CP}}^{\text{S2}})$. In Nash bargaining, such a point is called the \emph{starting point} \cite{game_concepts}. We denote it as $(\mathds{U}^{s}_{\text{ISP}},\mathds{U}^{s}_{\text{CP}})$. Then, according to the fairness concept of NBS, the fair profit distribution will be on the Pareto boundary and can be deduced by
\begin{equation}
\begin{array}{rl}
   \underset{\mathds{U}_{\text{ISP}},\mathds{U}_{\text{CP}}}{\text{maximize}}& (\mathds{U}_{\text{ISP}}-\mathds{U}^{s}_{\text{ISP}})(\mathds{U}_{\text{CP}}-U^{s}_{\text{CP}}),\\
    \text{subject to} & \mathds{U}_{\text{ISP}}+\mathds{U}_{\text{CP}}=\mathds{U}^{\text{S3}}_{\text{total}}.
\end{array}
\end{equation}
Here, NBS satisfies all the following four axioms \cite{game1,game_concepts,NBS}: (1) Invariant to equivalent utility representations; (2) Pareto optimality; (3) Independence of irrelevant alternatives; and (4) Symmetry. By solving the above optimization problem, we can obtain a fair profit distribution as follows:
\begin{equation}\label{nbs-re}
  \begin{array}{ll}
    \mathds{U}^{\text{S3}}_{\text{ISP}}=\mathds{U}^{s}_{\text{ISP}}+\frac{\mathds{U}^{\text{S3}}_{\text{total}}-\mathds{U}^{s}_{\text{ISP}}-\mathds{U}^{s}_{\text{CP}}}{2}, \\
    \mathds{U}^{\text{S3}}_{\text{CP}}=\mathds{U}^{s}_{\text{CP}}+\frac{\mathds{U}^{\text{S3}}_{\text{total}}-\mathds{U}^{s}_{\text{ISP}}-\mathds{U}^{s}_{\text{CP}}}{2}.
  \end{array}
\end{equation}
Then, the profit that PCP coalition should transfer to ISP coalition is $\mathcal{R}=\mathds{U}^{\text{S3}}_{\text{ISP}}-\mathds{U}^{\text{S3'}}_{\text{ISP}}=\mathds{U}^{\text{S3'}}_{\text{CP}}-\mathds{U}^{\text{S3}}_{\text{CP}}$.\\
\indent For different traffic profiles, we illustrate the improvement of $\mathds{U}^{\text{S3}}_{\text{ISP}}$ and $\mathds{U}^{\text{S3}}_{\text{CP}}$ compared with starting point as we have analyzed in Section~\ref{state-transmission-analysis}) in Fig.~\ref{improve-for-ab}. Fig.~\ref{improve-for-ab} shows that compared with the starting point, $\mathds{U}_{\text{ISP}}$ increases by more than $110\%$ and $\mathds{U}_{\text{CP}}$ increases by more than $70\%$.\\
\indent Specifically, for $\alpha=0.6$ and $\beta=0.3$, the Nash bargaining between ISP and PCP coalitions is illustrated in Fig.~\ref{example}, which shows that the corresponding starting point is $\left(\mathds{U}_{\text{ISP}}^{\text{S2}},\mathds{U}_{\text{CP}}^{\text{S2}}\right)=(3.5180,5.6450)$. According to Eq.(\ref{nbs-re}), we can obtain $(\mathds{U}^{\text{S3}}_{\text{ISP}},\mathds{U}^{\text{S3}}_{\text{CP}})=(8.6508,10.7778)$ as the final profit distribution. The profit that PCP coalition should assign to ISP coalition is  $\mathcal{R}=3.7449$. Compared with the starting point, $\mathds{U}_{\text{ISP}}$ increases by 145.90\% and $\mathds{U}_{\text{CP}}$ increases by 90.92\%. Thus, both ISP coalition and PCP coalition benefit much from this cooperation.

\section{Profit Distribution Within Each Coalition}
\label{sharing}

In this section, we will propose a mechanism to determine profit distribution within each coalition and discuss on its fairness and feasibility.

\subsection{Profit Distribution Mechanism}
\label{coalition-share}
To guarantee the stability of each coalition, the profit distribution mechanism should have the property of fairness. Before the mechanism, we give some definitions first.\\
\indent Suppose there are $m$ ISPs and $n$ PCPs. For the \mbox{$i$-th} PCP ($1\leq i\leq n$), we define two \emph{traffic matrices} as follows:
\begin{enumerate}
  \item $\mathbf{T}_i=\left(t^i_{j,k}\right)_{m\times m}$, where $t^i_{j,k}$ denotes the amount of the \mbox{$i$-th} PCP's traffic volume transmitted from the users in the \mbox{$j$-th} ISP's network to the users in the \mbox{$k$-th} ISP's network; and
  \item $\widetilde{\mathbf{T}}_i=\text{diag}(\tilde{t}^{i}_1,\tilde{t}^{i}_2,\cdots,\tilde{t}^{i}_m)$, where $\tilde{t}^{i}_j$ denotes the amount of the \mbox{$i$-th} PCP's traffic volume transmitted from its content servers to the users in the \mbox{$j$-th} ISP's network (Note that this part of uploading traffic will be charged by the corresponding ISP on the \mbox{$i$-th} PCP side).
\end{enumerate}
Thus, in PCP coalition, the amount of traffic volume caused by the \mbox{$i$-th} PCP accounts for a proportion
\begin{equation}\label{PCP-pro}
    \varphi_i=\frac{\left(\sum\limits_{1\leq j,k\leq m}{t^i_{j,k}}\right)+\left(\sum\limits_{j=1}^{m}{\tilde{t}^{i}_j}\right)}{v_{\text{p2p}}},
\end{equation}
\indent For ISP coalition, its two corresponding \emph{aggregated traffic matrices} are defined by
\begin{equation*}\label{isp-matrix}
    \mathbb{T}=\sum\limits_{i=1}^{n}{\mathbf{T}_i},
    \widetilde{\mathbb{T}}=\sum\limits_{i=1}^{n}{\widetilde{\mathbf{T}}_i}.
\end{equation*}
Suppose $\mathbb{T}=\left(t_{j,k}\right)_{m\times m}$ and $\widetilde{\mathbb{T}}=\text{diag}(\tilde{t}_1,\tilde{t}_2,\cdots,\tilde{t}_m)$. Then, in the \mbox{$l$-th} ISP's network (where $1\leq l\leq m$), the amount of P2P traffic volume caused by PCP coalition on users' side is
\begin{equation} \label{bbar}
    \varpi_l=\left(\sum\limits_{k=l}^{k=m}{t_{l,k}}\right)+\left(\sum\limits_{k=l}^{k=m}{t_{k,l}}\right)+\tilde{t}_l.
\end{equation}
In addition, in the \mbox{$l$-th} ISP's network, let $v_l$ and $b_l$ be the total traffic volume on users' side and the total bandwidth bought by all users using a flat price, respectively. It should be noted that $\sum\limits_{l=1}^{m}{b_l}=b_{\text{user}}^{\text{S0}}$. Then, in the \mbox{$l$-th} ISP's network, we can verify that the amount of background C/S traffic volume is $v_l-\varpi_l$ and \mbox{free-riding} P2P traffic volume is $v_l-b_l\cdot\xi_{\text{user}}$. According to the network model described in Section \ref{network-model}, clearly,
\begin{equation*}
    \sum\limits_{l=1}^{m}{\left[b_l\cdot\xi_{\text{user}}-(v_l-\varpi_l)\right]}=v^{\text{S0}}\cdot \alpha.
\end{equation*}
In addition, we can deduce that the \mbox{$l$-th} ISP's contribution to the free riding of P2P traffic accounts for a proportion
\begin{equation}\label{ISP-pro1}
    \psi_l=\frac{v_l-b_l\cdot\xi_{\text{user}}}{v_{\text{p2p}}\cdot(2-\beta)-v^{\text{S0}}\cdot \alpha}.
\end{equation}
\indent Consequently, we propose a fair and feasible profit distribution mechanism. For a given $\mathcal{R}$, the profit that the \mbox{$i$-th} PCP should assign to ISP coalition is $\mathcal{R}\cdot \varphi_i$ and the profit that ISP coalition should assign to the \mbox{$l$-th} ISP is $\mathcal{R}\cdot \psi_l$.\\
\indent Consider the example in the previous section. Suppose $\mathcal{M}_{\text{ISP}}=\left\{\text{ISP}_1,\text{ISP}_2,\text{ISP}_3\right\}$, $\mathcal{M}_{\text{PCP}}=\left\{\text{PCP}_1,\text{PCP}_2\right\}$ and
\begin{equation*}
  \begin{array}{rcl}
    \mathbf{T}_1&=&\left(\begin{array}{ccc}
 0.8255 &   1.6509  & 2.4764\\
 1.6509   & 1.6509   & 3.3019\\
 0.8255   & 1.6509  &  1.6509
 \end{array}\right),\\
   \mathbf{T}_2&=&\left(\begin{array}{ccc}
 2.4764  &  1.2382  &  3.7146\\
 1.2382  &  2.4764  &  1.2382\\
 4.9528  &  2.4764  &  3.7146
\end{array}\right),
  \end{array}
\end{equation*}
\begin{equation*}
  \begin{array}{rcl}
    \widetilde{\mathbf{T}}_1&=&\text{diag}(1.4151,2.1226,3.1840),\\
    \widetilde{\mathbf{T}}_2&=&\text{diag}(3.7146,2.6533,3.7146).
  \end{array}
\end{equation*}
 Thus, we can deduce that the profit PCP coalition should assign to $\text{ISP}_1$, $\text{ISP}_2$ and $\text{ISP}_3$ are 1.1598, 1.0812 and 1.5040, respectively.
\begin{figure}[h]
\centering
\subfigure[PCP]{
\includegraphics[width=2.2in]{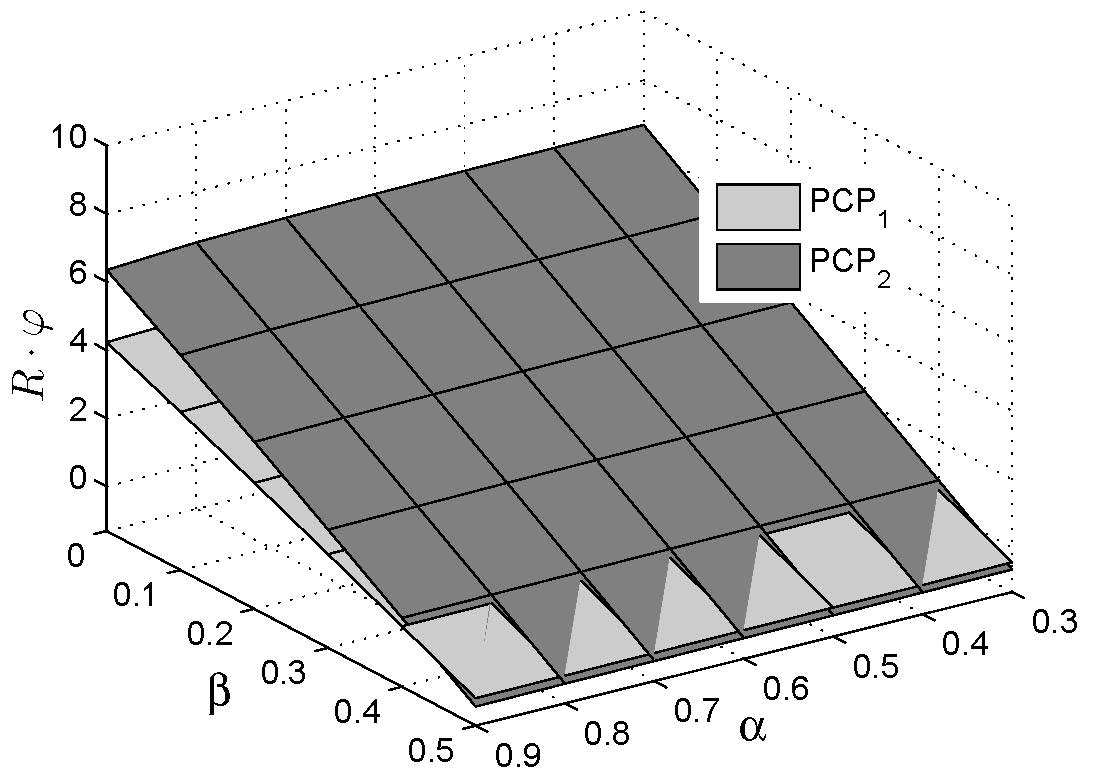}
}
\subfigure[ISP]{
\includegraphics[width=2.2in]{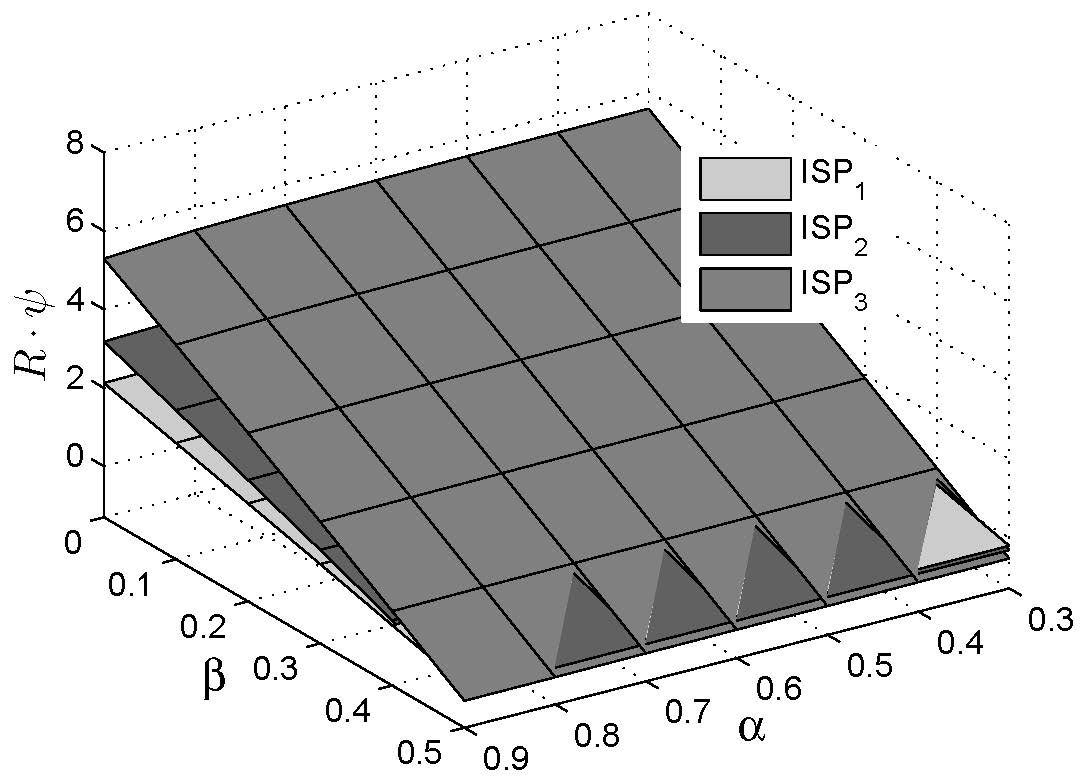}
}
\caption{PCP and ISP profit transfer with different $\alpha$ and $\beta$.}\label{inner-for-ab}
\end{figure}
Now, consider a more general example with $\left(\varphi_1,\varphi_2\right)=(0.4,0.6)$. We suppose the numbers of users in $\text{ISP}_1$, $\text{ISP}_2$ and $\text{ISP}_3$ are $N_1$, $N_2$ and $N_3$, respectively and the ratio of $N_1:N_2:N_3$ is $2:3:5$. Let $\mathbb{N}=N_1+N_2+N_3$. Moreover, suppose all users have the same preferences and behaviors. Then, the initially bought bandwidth is proportional to the number of users, $b_1:b_2:b_3=N_1:N_2:N_3=2:3:5$. And the requirements of background traffic are also proportional to the number of users, i.e., $v^{\text{S0}}_{\text{CS1}}:v^{\text{S0}}_{\text{CS2}}:v^{\text{S0}}_{\text{CS3}}=N_1:N_2:N_3$. Besides, we assume P2P applications use random peer selection scheme and the contents are distributed uniformly among users. For the \mbox{$i$-th} P2P application, suppose the average usage of each user is $\sigma_{i}$ ($i=1,2$). Then, we can obtain the traffic matrix of the \mbox{$i$-th} PCP as following
\begin{equation*}
    \mathbf{T}_i=\sigma_i\cdot(1-\beta)\cdot\left(\begin{array}{ccc}
  N_1\cdot\frac{N_1}{\mathbb{N}} & N_2\cdot\frac{N_1}{\mathbb{N}} & N_3\cdot\frac{N_1}{\mathbb{N}} \\
 N_1\cdot\frac{N_2}{\mathbb{N}} & N_2\cdot\frac{N_2}{\mathbb{N}}  & N_3\cdot\frac{N_2}{\mathbb{N}}\\
N_1\cdot\frac{N_3}{\mathbb{N}} & N_2\cdot\frac{N_3}{\mathbb{N}}  &  N_3\cdot\frac{N_3}{\mathbb{N}}
 \end{array}\right).
\end{equation*}
Moreover, based on our assumptions on users' behaviors and the above matrixes, we know that traffic provided by servers for each network are also proportional to $N_1:N_2:N_3$:
\begin{equation*}
    \widetilde{\mathbf{T}}_i=\text{diag}(\sigma_i\cdot N_1\cdot\beta, \sigma_i\cdot N_2\cdot\beta, \sigma_i\cdot N_3\cdot\beta).
\end{equation*}
Therefore, based on Eq.(\ref{ISP-pro1}), we can get the ratio of three ISPs' contribution weight as $\psi_1:\psi_2:\psi_3=N_1:N_2:N_3=2:3:5$. Fig.~\ref{inner-for-ab} illustrates the amount of profit transfer of each PCP and each ISP under different traffic profiles ($0.3\leq\alpha\leq 0.9$ and $0\leq\beta\leq 0.5$). We can see that with the increase of $\alpha$ or $\beta$, such profit transfer will decrease.

\subsection{Discussion}
\indent \textbf{\emph{Fairness}}. Fairness of this mechanism is guaranteed by two characteristics of $\varphi_i$ and $\psi_l$: (1) $\varphi_i$ increases with the total traffic volume of the \mbox{$i$-th} PCP; and (2) $\psi_l$ increases with the total traffic volume on users' side in the \mbox{$l$-th} ISP's network, but decreases with the total bandwidth bought by all users using a flat price in the \mbox{$l$-th} ISP's network. It can be easily verified that this profit distribution mechanism has the following properties: efficiency, symmetry and dummy player \cite{Revenue-sharing,Shapley1,shapley}.\\
\indent \textbf{\emph{Feasibility}}. This mechanism can be easily implemented because: (1) This mechanism is compatible with the traditional Internet economic settlement. Transit ISPs do not join this cooperation and the transit traffic still can be charged according to the old economic agreements between transit ISPs and eyeball ISPs; (2) All the information required by this profit distribution mechanism can be easily collected from ISPs and PCPs; (3) The calculation of this mechanism is easy.

\section{Related work}\label{related-work}
There are two categories of strategies for ISPs to address the problems caused by \mbox{P2P}. One belongs to engineering means, which includes (1) negative resistance against P2P by throttling, shaping, or blocking \cite{cat-mouse,P2Pfea} and (2) cooperation with PCPs to efficiently manage P2P traffic \cite{P2P_ISP2,P4P,taming,peer-sel,P2P-ISP}. The former runs counter to the development trend of P2P technology and may lead to PCPs' countermeasures, such as encryption and dynamic ports; while the latter may involve legal problems and privacy issues.\\
\indent Another category belongs to economic schemes. He \emph{et al}. \cite{survey} surveyed Internet pricing models and concluded that pricing acts as an important auxiliary to control network traffic. On this problem, one research direction is that ISPs change pricing models towards P2P users \cite{uplink-pricing,survey}. Two layers of relationships (\mbox{ISP-users} and \mbox{ISP-ISP}) are often studied based on \mbox{non-cooperative} game model \cite{game_concepts,game1}. And other researches giving insights to our work focus on the cooperative interactions among CPs, ISPs and users, which aim to find a multilateral satisfactory solution. For example, Hande \emph{et al.} \cite{CP-participation} proposed \mbox{CP-aided} flow pricing to optimize the utilities of ISP and CP. Misra \emph{et al.} \cite{fluid} proposed to fairly share profits based on Shapley value \cite{shapley} to stimulate \mbox{peer-assisted} services. Altman \emph{et al.} \cite{altman-bargaining,altman} studied the charging and revenue distribution between ISP and CP with different bargain power based on Nash bargaining \cite{NBS}.\\
\indent In this paper, we start from modeling \mbox{non-cooperative} game behaviors of CPs, ISPs and users with a practical utility optimization model. Then, we analyze a \mbox{two-player} {non-cooperative} dynamic game between ISPs and PCPs. Since cooperative game theory \cite{NBS,shapley} has been applied to many network fields and shows desirable properties in corresponding solutions \cite{Revenue-sharing,Shapley1,altman-bargaining,altman,TE-NBS}, we use \emph{Nash Bargaining Solution} \cite{NBS} to explore the benefits of cooperation and guarantee fair profit distribution. To our knowledge, this is the first work that systematically studies solutions for P2P caused unbalanced profit distribution and gives a feasible cooperative method to increase and fairly share profit.
\section{Conclusions}\label{conclude}
Under traditional Internet pricing mechanism, \mbox{free-riding} P2P traffic causes an unbalanced profit distribution between PCPs and ISPs, which will drive ISPs to take actions against P2P, finally impeding the wide adoption of P2P technology. Therefore we propose a new cooperative \mbox{profit-distribution} model based on the concept of Nash bargaining, in which ISPs and PCPs form two coalitions respectively and then cooperate to maximize their total profit. The fair profit distribution between the two coalitions is determined based on Nash Bargaining Solution (NBS). To guarantee the stability of each coalition, a fair mechanism for profit distribution within each coalition was designed. What is worth mentioning is that users obtain higher utilities within this cooperation-based model. As a result, such a cooperative \mbox{profit-distribution} method not only guarantees the fair profit distribution among network participators, but also improves the economic efficiency of the overall network system.\\
\indent Practical issues deserve further study before the adoption of this new cooperative \mbox{profit-distribution} method. We will study the supervision mechanisms between ISPs and PCPs and those within each coalition. Also, we will consider the inherent competitive relations among members within each coalition.
\bibliographystyle{IEEEtran}
\bibliography{IEEEabrv,coop}

\begin{thebibliography}{10}
\providecommand{\url}[1]{#1}
\csname url@samestyle\endcsname
\providecommand{\newblock}{\relax}
\providecommand{\bibinfo}[2]{#2}
\providecommand{\BIBentrySTDinterwordspacing}{\spaceskip=0pt\relax}
\providecommand{\BIBentryALTinterwordstretchfactor}{4}
\providecommand{\BIBentryALTinterwordspacing}{\spaceskip=\fontdimen2\font plus
\BIBentryALTinterwordstretchfactor\fontdimen3\font minus
  \fontdimen4\font\relax}
\providecommand{\BIBforeignlanguage}[2]{{%
\expandafter\ifx\csname l@#1\endcsname\relax
\typeout{** WARNING: IEEEtran.bst: No hyphenation pattern has been}%
\typeout{** loaded for the language `#1'. Using the pattern for}%
\typeout{** the default language instead.}%
\else
\language=\csname l@#1\endcsname
\fi
#2}}
\providecommand{\BIBdecl}{\relax}
\BIBdecl

\bibitem{P2P_VOD-design}
Y.~Huang, T.~Z.~J. Fu, D.-M. Chiu, J.~C.~S. Lui, and C.~Huang, ``{Challenges,
  Design and Analysis of a Large-scale {P2P-VoD} System},'' in \emph{Proc. ACM
  SIGCOMM'08}, Seattle, WA, Aug. 2008, pp. 375--388.

\bibitem{P2P-vod}
B.~Cheng, L.~Stein, H.~Jin, X.~Liao, and Z.~Zhang, ``{GridCast}: improving peer
  sharing for {P2P VoD},'' \emph{{ACM} Transactions on Multimedia Computing,
  Communications and Applications}, vol.~4, no.~4, pp. 1--31, 2008.

\bibitem{P2P-vod-design2}
K.~Suh, C.~Diot, J.~Kurose, L.~Massoulie, C.~Neumann, D.~F. Towsley, and
  M.~Varvello, ``Push-to-peer video-on-demand system: Design and evaluation,''
  \emph{{IEEE} J. Sel. Areas Commun.}, vol.~25, no.~9, pp. 1706--1716, 2007.

\bibitem{P2P_saving}
V.~Valancius, N.~Laoutaris, L.~Massoulie, C.~Diot, and P.~Rodriguez,
  ``{Greening the Internet with nano Data Centers},'' in \emph{Proc. {ACM
  CoNEXT}'09}, Rome, Italy, Dec. 2009.

\bibitem{P2P_saving2}
C.~Huang, J.~Li, and K.~W. Ross, ``{Can Internet video-on-demand be
  profitable?}'' in \emph{Proc. {ACM SIGCOMM}'07}, Kyoto, Japan, Aug. 2007.

\bibitem{Revenue-sharing}
R.~T.~B. Ma, V.~Misra, D.~ming Chiu, D.~Rubenstein, and J.~C.~S. Lui, ``{On
  Cooperative Settlement Between Content, Transit and Eyeball Internet Service
  Providers},'' in \emph{Proc. {ACM CoNEXT}'08}, Dec. 2008.

\bibitem{flat}
A.~M. Odlyzko, ``Internet pricing and the history of communications,''
  \emph{Computer Networks}, vol.~36, no.~5, pp. 493--517, 2001.

\bibitem{P2Pfea}
P.~Rodriguez, S.-M. Tan, and C.~Gkantsidis, ``{On the Feasibility of
  Commercial, Legal {P2P} Content Distribution},'' \emph{ACM SIGCOMM Computer
  Communication Review}, vol.~36, no.~1, pp. 75--78, 2006.

\bibitem{his_price2}
R.~Cocchi, S.~Shenker, D.~Estrin, and L.~Zhang, ``Pricing in computer networks:
  motivation, formulation, and example,'' \emph{{IEEE/ACM} Trans. Netw.},
  vol.~1, no.~6, pp. 614--627, Oct. 1993.

\bibitem{95th}
X.~Dimitropoulos, P.~Hurley, and A.~K.~M. Stoecklin, ``On the 95-percentile
  billing method,'' in \emph{Proc. {PAM'09}}, Seoul, South Korea, Apr. 2009.

\bibitem{cat-mouse}
\BIBentryALTinterwordspacing
``{Maximizing BitTorrent Speeds with uTorrent (Guide / Tutorial) Version
  1.161},'' Jul. 2010. [Online]. Available:
  \url{http://www.bootstrike.com/Articles/BitTorrentGuide/}
\BIBentrySTDinterwordspacing

\bibitem{P2P_ISP2}
G.~Shen, Y.~Wang, Y.~Xiong, B.~Y. Zhao, and Z.-L. Zhang, ``{HPTP}: Re-lieving
  the tension between {ISPs} and {P2P},'' in \emph{Proc. IPTPS'07}, Bellevue,
  WA, Feb. 2007.

\bibitem{P2P-ISP}
V.~Aggarwal, A.~Feldmann, and C.~Scheideler, ``Can {ISPs} and {P2P} users
  cooperate for improved performance?'' \emph{ACM SIGCOMM Computer
  Communication Review}, vol.~37, no.~3, pp. 29--40, 2007.

\bibitem{P4P}
H.~Xie, Y.~R. Yang, A.~Krishnamurthy, Y.~G. Liu, and A.~Silberschatz, ``{P4P}:
  provider portal for applications,'' in \emph{Proc. ACM SIGCOMM'08}, Seattle,
  WA, Aug. 2008.

\bibitem{CP-participation}
P.~Hande, M.~Chiang, R.~Calderbank, and S.~Rangan, ``Network pricing and rate
  allocation with content provider participation,'' in \emph{Proc. IEEE
  INFOCOM'09}, Rio de Janeiro, Brazil, 2009.

\bibitem{uplink-pricing}
Q.~Wang, D.~Chiu, and J.~C. Lui, ``{ISP Uplink Pricing in a Competitive
  Market},'' in \emph{Proc. ICT'08}, St. Petersburg, Russia, 2008.

\bibitem{survey}
\BIBentryALTinterwordspacing
H.~He, K.~Xu, and Y.~Liu, ``Internet resource pricing models, mechanisms, and
  methods,'' Submitted to Networking Science, Tech. Rep., Apr. 2011. [Online].
  Available: \url{http://arxiv.org/abs/1104.2005}
\BIBentrySTDinterwordspacing

\bibitem{NBS}
J.~F. Nash, ``The bargaining problem,'' \emph{Econometrica}, vol.~28, pp.
  155--162, 1950.

\bibitem{cp-price}
A.~Dhamdere and C.~Dovrolis, ``Can {ISPs} be profitable without violating
  ``network neutrality''?'' in \emph{Proc. {ACM NetEcon'08}}, Aug. 2008.

\bibitem{ISP-cost}
J.~K. MacKie-Mason and H.~R. Varian, ``{Pricing Congestible Network
  Resources},'' \emph{{IEEE} J. Sel. Areas Commun.}, vol.~13, no.~7, pp.
  1141--1149, Sep. 1995.

\bibitem{usage-ratio}
W.~B. Norton, ``{Video Internet: The Next Wave of Massive Disruption to the
  U.S. Peering Ecosystem},'' in \emph{Equinix white papers}, 2007.

\bibitem{game_concepts}
X.-R. Cao, H.-X. Shen, R.~Milito, and P.~Wirth, ``{Internet Pricing With a Game
  Theoretical Approach: Concepts and Examples},'' \emph{{IEEE/ACM} Trans.
  Netw.}, vol.~10, no.~2, pp. 208--216, Apr. 2002.

\bibitem{game1}
H.~Yaiche, R.~R. Mazumdar, and C.~Rosenberg, ``A game theoretic framework for
  bandwidth allocation and pricing in broadband networks,'' \emph{{IEEE/ACM}
  Trans. Netw.}, vol.~8, no.~5, pp. 667--678, Oct. 2000.

\bibitem{Shapley1}
R.~T.~B. Ma, D.~Chiu, J.~C.~S. Lui, V.~Misra, and D.~Rubenstein, ``{Internet
  economics: the use of Shapley value for ISP settlement},'' in \emph{Proc.
  {ACM CoNEXT}'07}, Dec. 2007.

\bibitem{shapley}
E.~Winter., \emph{{The Shapley Value}}.\hskip 1em plus 0.5em minus 0.4em\relax
  North-Holland: {in The Handbook of Game Theory. R. J. Aumann and S. Hart},
  2002.

\bibitem{taming}
D.~R. Choffnes and F.~Bustamante, ``Taming the torrent: A practical approach to
  reducing {cross-ISP} traffic in peer-to-peer systems,'' in \emph{Proc. ACM
  SIGCOMM'08}, Seattle, WA, Aug. 2008, pp. 363--374.

\bibitem{peer-sel}
R.~Bindal, P.~Cao, W.~Chan, J.~Medved, G.~Suwala, T.~Bates, and A.~Zhang,
  ``{Improving Traffic Locality in BitTorrent via Biased Neighbor Selection},''
  in \emph{Proc. ICDCS'06}, 2006, pp. 66--77.

\bibitem{fluid}
V.~Misra, S.~Ioannidis, A.~Chaintreau, and L.~Massoulie, ``Incentivizing
  peer-assisted services: a fluid shapley value approach,'' in \emph{Proc.
  SIGMETRICS'10}, 2010, pp. 215--226.

\bibitem{altman-bargaining}
E.~Altman, M.~K. Hanawal, and R.~Sundaresan, ``Nonneutral network and the role
  of bargaining power in side payments,'' in \emph{the 4th Workshop on Network
  Control and Optimization (NETCOOP)'10}, Ghent, Belgium, Nov. 2010, pp.
  66--73.

\bibitem{altman}
\BIBentryALTinterwordspacing
E.~Altman, A.~Legout, and Y.~Xu, ``Network non-neutrality debate: An economic
  analysis,'' CoRR abs/1012.5862:, 2010. [Online]. Available:
  \url{http://arxiv.org/abs/1012.5862}
\BIBentrySTDinterwordspacing

\bibitem{TE-NBS}
W.~Jiang, R.~Zhang-Shen, J.~Rexford, and M.~Chiang, ``{Cooperative Content
  Distribution and Traffic Engineering in an ISP Network},'' in \emph{Proc.
  SIGMETRICS/Performance'09}, Seattle, WA, 2009.

\end{thebibliography}
\end{document}